\documentclass[12pt]{article}
\usepackage{amssymb, amsmath,cite}    
\usepackage{amsthm}
\usepackage{xcolor}
\usepackage[labelfont={color=black,footnotesize},textfont={color=black,footnotesize,it,up}]{caption} 
\usepackage{url}
\usepackage[titletoc,toc,title]{appendix}
\theoremstyle{definition}

\usepackage{subfig,tikz}
\usetikzlibrary{decorations.pathmorphing}
\usepackage{graphics}

\newcommand{\Tr}{\,{\rm Tr}}

\newcommand{\td}{\text{d}}
\def\be{\begin{equation}}
\def\ee{\end{equation}}
\def\bea{\begin{eqnarray}}
\def\eea{\end{eqnarray}}

\title{\bf{On a  mass functional for initial data in $4+1$ dimensional spacetime}}

\author{Aghil Alaee\footnote{aak818@mun.ca } \,\,and  Hari K. Kunduri\footnote{hkkunduri@mun.ca } \\ \\
\small \sl $^a$ Department of Mathematics and Statistics, \\  \small \sl Memorial University of Newfoundland \\ \small \sl St John's NL A1C 4P5, Canada}

\date{}

\begin{document}

\maketitle






\begin{abstract}
We consider a broad class of asymptotically flat, maximal initial data sets satisfying the vacuum constraint equations, admitting two commuting rotational symmetries.  We construct a mass functional for `$t-\phi^i$' symmetric data which evaluates to the ADM mass.  We then show that $\mathbb{R} \times U(1)^2$-invariant solutions of the vacuum Einstein equations are critical points of this functional amongst this class of data.  We demonstrate positivity of this functional for a class of rod structures which include the Myers-Perry initial data. The construction is a natural extension of Dain's mass functional to $D=5$, although several new features arise. 
\end{abstract}

\section{Introduction} 
Dain has established the remarkable inequality $m \geq \sqrt{|J|}$, for complete asymptotically flat, axisymmetric maximal initial data $(\Sigma, h, K)$ of the vacuum Einstein equations in four dimensions \cite{dain2006proof,dain2008proof,dain2008inequality,dain2012geometric}.  Here $m$ is the ADM mass of the Riemannian manifold and $J$ is the conserved angular momenta, defined from the existence of the $U(1)$ isometry.  This result was subsequently strengthened to a more general class of metrics, multiple asymptotic ends and weaker asymptotic fall-off conditions \cite{Chrusciel2009}. Dain's equality is saturated if and only if $(\Sigma, h, K)$ is that of constant-Boyer-Lindquist time hypersurface of the extreme Kerr black hole. 

An important step in the proof of this inequality was the construction of a well-defined mass functional $\mathcal{M}$, defined for $t-\phi$ symmetric, asymptotically flat maximal initial data. $\mathcal{M}=\mathcal{M}(v,Y)$ depends on two scalar functions $v$ and $Y$ which can be shown to fully specify the initial data set.   The proof shows that $m = \mathcal{M}(v,Y)$ for $t-\phi$ symmetric maximal initial data, and that $m \geq \mathcal{M}$ for arbitrary axisymmetric maximal data.  $\mathcal{M}(v,Y)$ can be shown to be positive-definite and the unique minimizer is extreme Kerr, completing the elegant argument. 

It is natural to expect an analogous inequality would hold in $D=5$ dimensions, under suitable restrictions on the initial data. The situation is particularly interesting as there are potentially two candidates for minimizers: extreme Myers-Perry black holes with $S^3$ horizon topology \cite{myers1986black}, and extreme black rings with $S^1 \times S^2$ horizon topology.  The masses of these solutions satisfy
\begin{eqnarray}
M^3 &=& \frac{27\pi}{32}\left(|J_1| + |J_2|\right)^2 \qquad \textrm{(Myers-Perry)} \\ M^3 &=& \frac{27\pi}{4} |J_1| (|J_2| - |J_1|) \qquad \textrm{(black ring)}
\end{eqnarray} where $J_i$ are conserved angular momenta computed in terms of Komar integrals. Of course it is not manifestly clear how an expression which is derived from the ADM mass (i.e. evaluated at spatial infinity) would capture information on the topology of the horizon - indeed, at the level of the initial data,  the horizon is a minimal surface in the interior.  It is worth noting that another, related class of geometric inequalities relating the \emph{area} of marginally outer trapped surfaces to the angular momenta (and charge) have also been established in three spatial dimensions\cite{dain2011area,jaramillo2011black}.  Once again the geometries which uniquely saturate the bound were the horizon geometries corresponding to the extreme Kerr geometry.  Recently, Hollands has derived an area-angular momenta inequality in general dimension $D$, for spaces admitting a $U(1)^{D-3}$ action\footnote{We write $U(1)^s \equiv U(1)\times U(1) \times \ldots \times U(1)$ with $s$ $U(1)$ factors.} as isometries\cite{hollands2012horizon}. In this case, the inequality depends on the topology of  the marginally outermost trapped surface. 

As a first step towards establishing a mass-angular momenta inequality in five dimensions, in this work we investigate a generalization of Dain's mass functional $M(v,Y)$ to $D>4$ for maximal spatial slices of five-dimensional vacuum spacetimes with $U(1)^2$ isometry.  Note that most of the local analysis works equally well for $D-$dimensional spacetimes with $U(1)^{D-3}$ isometry.  However, such spacetimes could only be asymptotically flat for $D=5$ (there might be a useful extension to spaces that are asymptotically Kaluza-Klein).  Hence we will focus on this case, although it will be sometimes convenient to leave $D$ as a free parameter.

Our first goal  is to construct a positive-definite functional which evaluates to the mass for a broad class of asymptotically flat, maximal initial data with `$t-\phi^i$'  symmetry ($i=1\ldots D-3$).  Such data can be thought of as data that is `stationary at a moment in time'. In particular, it allows us to specify the extrinsic curvature in terms of $D-3$ twist potentials, using the construction of transverse traceless tensors given in \cite{alaee2014small}.  The functional is defined over functions on the orbit space $\mathcal{B} \equiv \Sigma / U(1)^{D-3}$ and depends on a matrix $\lambda'_{ij}$ specifying the metric on surfaces of transitivity of the $U(1)^{D-3}$ action (often called the `Gram' matrix) and $D-3$ twist potentials. Setting $D=5$, this amounts to five independent functions.

Carter has established a variational formulation of the stationary, axisymmetric vacuum Einstein equations \cite{carter1973black}.  Our main result is to demonstrate the mass functional we have defined, when integrated over an appropriate domain. is the same as Carter's functional up to (divergent) boundary terms.  Therefore $\mathbb{R} \times U(1)^2$-invariant vacuum solutions arise as critical points of the mass amongst all asymptotically flat, $t-\phi^i$ symmetric initial data (see Bardeen's result \cite{bardeen1970variational} for the 3+1-dimensional case).  In this sense our proposed functional is an extension of Dain's functional $\mathcal{M}(v,Y)$, which also has this property.  However, there are a number of important differences.  As we will elaborate,  our functional contains boundary terms which encode the `rod structure' of the initial data.  In particular, the initial data may contain 2-cycles (`bubbles') which also contribute to the mass.  The rod structure plays an important role in the black hole uniqueness theorem \cite{hollands2008uniqueness, Figueras2010}  in five-dimensional vacuum gravity. In the case of stationary black holes containing additional 2-cycles, the usual laws of black hole mechanics have been shown to be modified \cite{Kunduri:2013vka}.  More recently, an explicit example of a black hole spacetime containing an 2 cycle in the exterior region was constructed in supergravity \cite{Kunduri:2014iga}.

This paper is organized as follows.  In Section 2 we introduce a broad class of $U(1)^{2}$-invariant  maximal initial data $(\Sigma,h_{ab} K_{ab})$ for the vacuum Einstein equations and discuss the particular case of $t-\phi^i$-symmetric data, which allows us to construct a general class of transverse-traceless tensors in the geometry. The resulting data is parameterized by functions on the orbit space and we discuss in detail their various boundary and asymptotic conditions that we must impose. In Section 3 we introduce a functional defined for asymptotically flat initial data which evaluates to the ADM mass and discuss some of its properties.  Section 4 investigates the relationship of this functional to Carter's variational formulation for stationary, axisymmetric vacuum solutions.  We conclude with an argument that demonstrates positivity of this functional for a particular class of rod structure which includes Myers-Perry black hole initial data. 
\section{Initial data with rotational isometries}
\subsection{Conformal Data}
Consider a stationary vacuum solution of the Einstein equations with $U(1)^{D-3}$ isometry group. It is a well known result \cite{harmark2004stationary,hollands2008uniqueness} that Forebenius' theorem and the vacuum equations imply orthogonal transitivity of the $\mathbb{R} \times U(1)^{D-3}$ action, and the metric can be written in the form
\begin{equation}
g = G_{\alpha \beta} \td\xi^\alpha \td\xi^\beta + g_{AB}\td x^A \td x^B
\end{equation} where $\td/ \td \xi^{\alpha}$ generate the isometry group ($\alpha,\beta=0,1,2$)  and $x^A$ are coordinates on the two-dimensional surfaces orthogonal to the surfaces of transitivity. We may write this explicitly as
\begin{equation}\label{weylspt}
g = -H \td t^2 + \frac{\lambda'_{ij}}{H^{1/N}}(\td\phi^i - w^i dt)(\td\phi^j - w^j \td t) + e^{2\nu} (\td\rho^2 + \td z^2)
\end{equation} where $N=D-1$ and $\rho^2 =\det \lambda'$ is harmonic on the spacetime orbit space $\tilde{\mathcal{B}} \equiv M / U(1)^{D-3}$ and $\td z$ is the harmonic conjugate of $\td\rho$.  Further details on the functions $\lambda_{ij}$ and one-forms $\omega^i$, and an analysis of $\tilde{\mathcal{B}}$, are given in \cite{Figueras2010,hollands2008uniqueness}.  Note that constant time slices in this spacetime have induced metric $h = H^{-1/N} \tilde{h}$ where
\begin{equation} \label{idmet}
\tilde{h} = \lambda'_{ij} \td\phi^i \td\phi^j + e^{2U} (\td\rho^2 + \td z^2)
\end{equation} where $e^{2U} = e^{2\nu} H^{1/N}$.  If we consider five-dimensional asymptotically flat black hole solutions, it is known that in an appropriate coordinate system, the spacetime metric takes the form \cite{hollands2008uniqueness}
\begin{eqnarray}
\td s^2&=&-\left(1-\frac{\mu}{R^2}+\mathcal{O}(R^{-2})\right)\td t^2+\left(\frac{2\mu a_1(R^2+a_1^2)}{R^4}\sin^2\theta+\mathcal{O}(R^{-3})\right)\td t\td\phi_1\nonumber\\
&+&\left(\frac{2\mu a_2(R^2+a_2^2)}{R^4}\cos^2\theta+\mathcal{O}(R^{-3})\right)\td t\td\phi_2+\left(1+\frac{\mu}{2R^2}+\mathcal{O}(R^{-3})\right)\nonumber\\
&\times&\Bigg[\frac{R^2+a_1^2\cos^2\theta+a_2^2\sin^2\theta}{(R^2+a_1^2)(R^2+a_2^2)}R^2\td R^2+(R^2+a_1^2\cos^2\theta+a_2^2\sin^2\theta)\td\theta^2\nonumber\\
&+&(R^2+a_1^2)\sin^2\theta\td\phi_1+(R^2+a_2^2)\cos^2\theta\td\phi_2\Bigg]
\end{eqnarray} where $R \to \infty$ corresponds to spatial infinity and $(\mu,a_i)$ are parameters related to the mass $M$ and angular momenta $J_i$ of the black hole respectively.  It is clear that $t=$constant slices in the above asymptotic geometry can be written $h = \Phi^2 \tilde{h}$, where $\tilde{h}$ has vanishing ADM mass. 

We now focus attention on a general class of vacuum initial data. It is important to note that the results discussed in this work apply to spacetimes which will evolve from this data. In particular, the evolution need not be stationary, and so the results apply to dynamical spacetimes. We will consider solutions of the vacuum constraint equations in $N$ space dimensions (Latin indices $a,b=1...N$)
\begin{eqnarray}\label{eq:constraints}
R_h - K^{ab}K_{ab} + (\Tr_h K)^2 &=& 0 \nonumber \\ \nabla^b\left(K_{ab} - \Tr_h K h_{ab}\right) &=& 0
\end{eqnarray} where $(\Sigma,h_{ab},K_{ab})$ refer to an asymptotically flat Riemannian manifold $(\Sigma,h_{ab})$ with second fundamental form $K_{ab}$ in spacetime. This initial data set is assumed to be \emph{maximal} ($\Tr_h K =0$) and invariant under a $U(1)^{N-2}$ isometry group generated by commuting Killing vector fields $m_i$, that is
\begin{equation}\label{axidata}
\mathcal{L}_{m_i} h_{ab} = 0\, , \qquad \mathcal{L}_{m_i} K_{ab} = 0
\end{equation}  Motivated by the above asymptotic geometry of black hole slices, we will focus on the case $N=4$ and $h_{ab}$ is \emph{conformal} to a $U(1)^{2}$ invariant metric $\tilde{h}$ of the form \eqref{idmet} with vanishing ADM mass.   This encompasses a broad class of initial data (we have checked this explicitly for the initial data for Myers-Perry black holes and the extreme doubly spinning black ring) . For generic initial data, of course, one need not have orthogonal transitivity of the $U(1)^{N-2}$ action, and, indeed, $\rho = \sqrt{\det \lambda'}$ need not be harmonic and so the two-dimensional metric will not take the conformally flat form above (i.e. $h_{\rho z} \neq 0$). We expect, however,  that these restrictions can be removed (see e.g. \cite{chrusciel2009mass}).

By Froebenius'  theorem the two-dimensional subspace of the tangent space at each point which are spanned by vectors orthogonal to $m_1$ and $m_2$ are integrable (tangent to two-dimensional surfaces) if and only if
\begin{gather}
\nabla_{[a}\eta_{b]}=l_{[a}\eta_{b]}+s_{[a}\gamma_{b]}\label{12}\\
\nabla_{[a}\gamma_{b]}=p_{[a}\eta_{b]}+q_{[a}\gamma_{b]}\label{13}
\end{gather} where we have set $\eta^a \equiv m_1^a$, $\gamma^a \equiv m_2^a$ and $\eta = \eta_a \eta^a$ and $\gamma = \gamma_a \gamma^a$, and $l_a$, $s_a$, $p_a$, and  $q_a$ are arbitrary 1-forms.  It is straightforward to verify that these imply the following identities:
\begin{gather}\label{hyp}
\begin{aligned}
\left(\det\lambda'\right)\nabla_a\eta_b&=\gamma\eta_{[b}\nabla_{a]}\eta-2L\gamma^c\eta_{[b}\nabla_{a]}\eta_{c}-L\gamma_{[b}\nabla_{a]}\eta+2\eta\gamma^c\gamma_{[b}\nabla_{a]}\eta_{c}\\
\left(\det\lambda'\right)\nabla_a\gamma_b&=-L\eta_{[b}\nabla_{a]}\gamma+2\gamma\gamma^c\eta_{[b}\nabla_{a]}\gamma_{c}+\eta\gamma_{[b}\nabla_{a]}\gamma-2L\gamma^c\gamma_{[b}\nabla_{a]}\gamma_{c}
\end{aligned}
\end{gather}  where $L = \eta_a \gamma^a$.
We will use these shortly. 

The Ricci tensor for the metric \eqref{idmet} is straightforward to compute \cite{harmark2004stationary}. We will divide the indices $a,b$ along $A,B=1 \ldots 2$ and $i,j=1 \ldots N-2$.   Our main interest is the scalar curvature. Since $\lambda'^{ij}R_{ij} =0$  we have
\begin{equation}\label{scalarcurvature}
\tilde{R} = g^{AB}R_{AB}  =  e^{-2U}\left[ -2 \nabla^2 U + \frac{1}{\rho^2}  - \frac{1}{4} \lambda'^{ik}\nabla_A \lambda'_{kj}  \lambda'^{jm} \nabla_B \lambda'_{mi} \delta^{AB}\right]
\end{equation} where $\nabla$ refers to the flat partial derivative operator $\partial_A$. We will also denote by $\cdot$ the scalar product with respect to the flat metric $\delta_{AB}$. The last term can also be written in the compact form
\begin{equation}
-\frac{1}{4} \Tr \left[ (\lambda'^{-1} \td\lambda')^2\right]
\end{equation}  If $N=3$, the final two terms in \eqref{scalarcurvature} cancel, and $\tilde{R}$ takes a particularly simple form.  This fact is crucial to establish positive-definiteness of the mass functional for three-dimensional initial data. 

We consider solutions $(\Sigma,h_{ab},K_{ab})$ of \eqref{eq:constraints}  expressed by the conformal rescaling
\begin{equation}\label{conformaldata}
h_{ab} = \Phi^2 \tilde{h}_{ab} \, ,\qquad K_{ab} = \Phi^{-2} \tilde{K}_{ab}
\end{equation} in terms of which the constraint equations for maximal slices become (note $\Tr_{\tilde{h}} \tilde K =0)$
\begin{gather}
\Delta_{\tilde{h}}\Phi-\frac{1}{6} R_{\tilde h}\Phi  +\frac{1}{6}\tilde{K}_{ab}\tilde{K}^{ab}\Phi^{-5}=0.\label{eq:Lich}\\
\tilde{\nabla}_b\tilde{K}^{ab}=0. \label{eq:divK}
\end{gather} The Lichnerowicz equation  \eqref{eq:Lich} is a second-order non-linear elliptic PDE for the conformal factor $\Phi$ on a fixed Riemannian manifold $(\Sigma,\tilde{h}_{ab})$ with a given symmetric, traceless, divergenceless rank two tensor field $\tilde{K}_{ab}$. The existence and uniqueness of solutions of \eqref{eq:Lich} is guaranteed by the results of \cite{choquet2000einstein} (see Section VIII) under suitable regularity conditions, i.e. $\tilde{K}_{ab}\tilde{K}^{ab}$ belongs to a particular weighted Sobolev space and $(\Sigma,\tilde{h})$ is in the positive Yamabe class. Clearly, the latter condition is true, because $\tilde{h}$ is conformal to $h$, which must have positive scalar curvature.  The former condition is an additional condition we impose on the data. 

We now focus attention on a class of axisymmetric initial data sets $(\Sigma,\tilde{h}_{ab}, \tilde{K}_{ab})$ for which the extrinsic curvature can be specified completely from scalar potentials. The construction of transverse, traceless tensors under these conditions is given in \cite{alaee2014small} and we will only briefly review it here.  Let $\phi^i$ be angular coordinates adapted to the commuting Killing fields $m_i$.  Following \cite{figueras2011black} we define an initial data set to be $t-\phi^i$-\emph{symmetric} if (1) $\partial / \partial \phi^i$ generate a  $U(1)^2$ isometry  and (2) $\phi^i \to -\phi^i$ is a diffeomorphism that preserves $\tilde{h}$ but reverses the sign of $\tilde{K}_{ab}$.  Condition (1) is obviously trivially satisfied by construction. In terms of the Weyl coordinate system used above, condition (2) implies $K_{ij}=K_{AB}=0$ (initial data with this property arise naturally within the context of slices of stationary, axisymmetric black holes for $N=3,4$).   Note that $t-\phi^i$ symmetry implies that the $U(1)^2$ action is orthogonally transitive, i.e. the identities \eqref{hyp} hold. 

As a consequence of this symmetry, $K_{ab}$ is automatically traceless. Using the divergenceless condition and the property $\Sigma$ is simply connected \cite{alaee2014small}, we can express $\tilde{K}_{ab}$ in a compact form.  Define two scalar potentials $Y_i$ and one-forms
\begin{equation}
S^i = \frac{1}{2 \det \lambda'} i_{m_1} i_{m_2} \star \td Y^i
\end{equation}  Note $\td \star S^i = 0$. Then an \emph{arbitrary} divergenceless $t-\phi^i$-symmetric extrinsic curvature can be expressed as \cite{alaee2014small}
\begin{equation}\label{exttphi}
\tilde{K}_{ab}=\frac{2}{\det\lambda'}\Bigg[\left(\lambda'_{22} S^1{}_{(a}m_1{}_{b)}-\lambda'_{12}S^2{}_{(a}m_1{}_{b)}\right)+\left(\lambda'_{11} S^2{}_{(a}m_2{}_{b)}-\lambda'_{12}S^1{}_{(a}m_2{}_{b)}\right)\Bigg].
\end{equation}  The vanishing of the trace of \eqref{exttphi} is obvious since $S^i$ and the $m_i$ are orthogonal. The divergenceless condition is more invovled and requires the use of the identities \eqref{hyp}.  Hence for $t-\phi^i$ symmetric initial data, the extrinsic curvature is completely characterized by the scalar potentials $Y^i$ as well as the metric functions $\lambda'_{ij}$. One can show \cite{alaee2014small} that these potentials are simply the pull-backs of the spacetime twist potentials defined in the usual way, i.e. $\td Y^i = \star_5(m_1 \wedge m_2 \wedge \td m_i)$.

Further, it is useful to note that the full contraction of this tensor is
\begin{eqnarray}\label{tildeK^2}
\tilde{K}_{ab}\tilde{K}^{ab}
&=&e^{-2U}\frac{\text{Tr}\left(\lambda'^{-1}\td Y \cdot \td Y^t\right)}{2\det\lambda'}
\end{eqnarray}
where for simplicity we use the notation $\td Y=\left(\td Y^1,\td Y^2\right)^t$ to define a  column vector.  If we consider the extrinsic curvature $\bar{K}_{ab}$ of a $U(1)^2$-invariant, non $t-\phi^i$-symmetric initial data set, one can show that 
\begin{equation}\label{Kaxi}
\bar{K}_{ab}\bar{K}^{ab} =  \tilde{K}_{ab} \tilde{K}^{ab} + K_{AB}K_{CD} g^{AC}g^{BD} + K_{ij}K_{kl}\lambda'^{ik}\lambda'^{jl} \geq \tilde{K}_{ab} \tilde{K}^{ab}
\end{equation} where $g_{AB} = e^{2U}\delta_{AB}$. In particular there is equality if and only if $t-\phi^i$-symmetry holds.  

In summary, we are considering the class of $U(1)^2$-invariant maximal initial data sets $(h_{ab},K_{ab})$ which can be globally represented by the form \eqref{conformaldata} satisfying \eqref{axidata}. The conformal metric \eqref{idmet} has vanishing ADM mass and is specified by the functions $U$ and $\lambda'_{ij}$.  Finally, if we impose in addition that the initial data be $t-\phi^i$ symmetric, then the extrinsic curvature is fully characterized by specifying in addition to the other data, two twist potentials $Y^i$.   
\subsection{Geometry of $\Sigma$}
To conclude this section we discuss general properties of the manifold $(\Sigma,h)$ and its $U(1)^2$ action.  Assume $\Sigma$ is complete, oriented, simply connected, with possibly multiple asymptotic ends, each of which is asymptotically flat or asymptotically cylindrical.  There is always at least one end of the former type. As a simple example, the $t=0$ maximal initial data slice of the Schwarzschild spacetime has the topology $\mathbb{R} \times S^3$, which has two asymptotically flat ends.  Asymptotically cylindrical ends (the geometry approaches a product metric on $\mathbb{R} \times N$ where $N$ is a closed 3-manifold of positive Yamabe type)  arise in the context of initial data for extreme black holes.  For concreteness, we mainly focus in this paper on the situation where $\Sigma$ has two asymptotic ends.

First, note that Orlik and Raymond have classified closed, simply connected oriented smooth manifolds admitting a torus $U(1)^2$ action with no non-trivial discrete isotropy subgroups \cite{OrlikRaymond}.  They show that such manifolds must have the topology of connected sums of copies of $S^2 \times S^2$, $\mathbb{CP}^2$, and $\overline{\mathbb{CP}}^2$ (note that taking the connected sum with $S^4$ is the identity operation). One may obtain asymptotically flat ends by removing points, or equivalently,  taking the connected sum with $\mathbb{R}^4$. For example, the  topology of the maximal slice of Schwarzschild discussed above can be obtained simply by removing two points from $S^4$. 

If we restrict to manifolds which admit a spin structure, the above decomposition will only contain copies of $S^2 \times S^2$.  Holland, Hollands, and Ishibashi have shown \cite{hollands2011further} the domain of outer communications for a stationary, asymptotically flat non-extreme black hole with spatial cross sections of the horizon $H$, satisfying the dominant energy condition with $\mathbb{R} \times U(1)$ isometry, has topology $\mathbb{R} \times \Sigma_0$ with
\begin{equation}\label{HHI}
\Sigma_0 \cong (\mathbb{R}^4 \# n (S^2 \times S^2)) \setminus B
\end{equation} where $B$ is a 4 manifold with closure $\bar{B}$ such that $\partial \bar{B} = H$ and we have again assumed the spacetime admits a spin structure.  For the case of the Myers-Perry black hole and the non-extreme black ring, the authors and Martinez-Pedroza have shown \cite{0264-9381-31-5-055004} that  $\Sigma_0 \cong [0,1)\times S^3$ and $\Sigma_0 \cong (S^2 \times D^2 )\# \mathbb{R}^4$ respectively (here $D^2$ refers to a closed two-disc).  Note that $\Sigma_0$ is asymptotically flat, simply connected,  and has an inner boundary $\partial \Sigma_0 = -H$. 

Given a compact manifold $M$, one may obtain a complete manifold $M'$ by \emph{doubling} $M$. The double of $M$ is the quotient space of $M \sqcup M$ obtained by identifying each point in the boundary $\partial M$ of the first copy of $M$ with the corresponding point in the boundary of the second copy \cite{Lee_smooth_manifolds}.  $M'$ is a smooth manifold without boundary. 
In the present case one can compactify $\Sigma_0$ by glueing in a closed 4-ball $D^4_\infty$, and then apply the doubling procedure to obtain a closed manifold. Finally, one may remove two (or more) points to obtain a complete manifold with asymptotically flat ends.  For example, a complete initial data slice for the non-extreme black ring can be obtained by doubling $\Sigma_0 \cong S^2 \times D^2 \#\mathbb{R}^4$ to yield\footnote{We thank G Galloway for a discussion on this point.}  $\Sigma \cong \mathbb{R}^4 \# (S^2 \times S^2) \# \mathbb{R}^4$. Note this space has two asymptotically flat ends with `boundaries at infinity' $S^3$. Complete initial data for the Myers-Perry black hole obtained by the analogous doubling procedure simply yields $\mathbb{R} \times S^3$. 

Examples of $\Sigma$ with asymptotically cylindrical ends can be obtained from the spatial slices of extreme black holes. In this case, spatial slices of the domain of outer communications are already complete (in contrast to the non-extreme case) because the horizon is an infinite proper distance away. The topology of $\Sigma$ can then be found from \eqref{HHI} where we now replace $B$ with its closure $\bar{B}$ to obtain a manifold without boundary.  For example,  for the extreme Myers-Perry black hole once again we have $\Sigma \cong \mathbb{R} \times S^3$ while for the extreme black ring, $\Sigma \cong (S^2 \times B^2) \# \mathbb{R}^4$. In the latter case the cylindrical end has $N \cong S^1 \times S^2$.  It is worth emphasizing that the topologies of a spatial slice of the non-extreme and extreme rings are different, in contrast with the situation for Myers-Perry, which behaves in the same way as the Kerr black hole in four dimensions. 

The initial data is best characterized in terms of the two-dimensional orbit space $\mathcal{B} \equiv \Sigma / U(1)^2$.  In general the action will have fixed points and so $\mathcal{B}$ may potentially have singularities. We assume there are no points with a discrete isotropy subgroup.  A careful analysis of the orbit space is performed in  \cite{hollands2008uniqueness} and we will summarize the results.   The projection $\pi: \Sigma \to \mathcal{B}$ defines a $U(1)^2$-principal bundle over the open subsets of $\mathcal{B}$ and we choose the bundle to have a flat connection (this is reflected by the fact $h_{\rho i} = h_{z i}=0$ for $t-\phi^i$-symmetric data). At the fixed points, some linear combination of the Killing fields $m_i$ must vanish, and the matrix $\gamma'_{ij}$ will be singular, so $\rho =0$. Note that the function $\rho$ is harmonic  with respect to the Laplacian of the orbit space metric $e^{2U}(\td \rho^2 + \td z^2)$ and the boundary of $\mathcal{B}$ is defined by $\rho =0$.   Asymptotic flatness of $\Sigma$ implies that $\rho$ must approach the corresponding quantity in Euclidean space outside a compact set, which is simply the radial coordinate $r$ in standard spherical coordinates in $\mathbb{R}^4$.  Hence $0 < \rho < \infty$ in the interior of $\mathcal{B}$ by the maximum principle. 

Furthermore, as $\Sigma$ is simply connected, connectedness of the $U(1)^2$ isometry group implies via standard homotopy arguments that $\mathcal{B}$ is simply connected  \cite{hollands2008uniqueness}.  The analysis of  \cite{hollands2008uniqueness} establishes  that $\mathcal{B}$ is a non-compact manifold with boundaries and corners, i.e. locally modelled on $\mathbb{R} \times \mathbb{R}$ at interior points, $\mathbb{R}^+ \times \mathbb{R}$ (neighbourhoods of one-dimensional boundary segments) and $\mathbb{R}^+ \times \mathbb{R}^+$ (neighbourhoods of corners).  As $\mathcal{B}$ is an oriented, simply connected analytic manifold with boundaries and corners, by the Riemann mapping theorem it can be analytically mapped to the upper complex plane, where we take $\rho \geq 0$ to lie on the non-negative imaginary axis and $z \in \mathbb{R}$ on the real axis. 

In the interior of $\mathcal{B}$, the matrix $\lambda'_{ij}$ has rank $2$, whereas on the boundary $\partial{\mathcal{B}} \cong \mathbb{R}$ and the conners it has rank $1$ and rank $0$ respectively.  $\partial{\mathcal{B}}$ consists of  finite spatial intervals (`rods'), and two semi-infinite intervals \cite{hollands2008uniqueness}.  To each of these intervals we associate a pair of integers $(p,q)$. These represent co-dimension 2 surfaces upon which an integer linear combination $p m_1 + q m_2$ of the rotational Killing fields vanish, and  the two rotation axes which extend to spatial infinity, respectively. For simplicity we choose our basis so that $(1,0)$ and $(0,1)$ vanish on the semi-infinite rods.   As we discuss in more detail below, the finite-length rods correspond to 2-cycles.  Thus in the absence of any additional  asymptotic ends this situation represents initial data on $\mathbb{R}^4$ with `bubbles', and is qualitatively different to the situation in $\mathbb{R}^3$, where topological censorship implies a  trivial second homology group. 
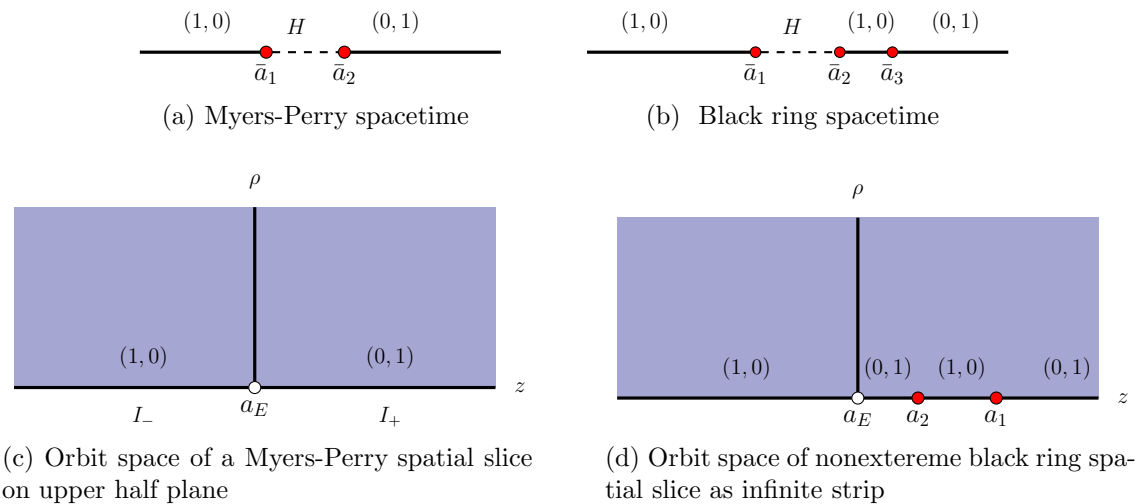
\begin{figure}[h]
\centering
\subfloat[{Myers-Perry spacetime}]{
\begin{tikzpicture}[scale=.8, every node/.style={scale=0.7}]
\draw[very thick](-3,0)--(-1,0)node[black,left=1cm,above=.2cm]{$(1,0)$};
\draw[thick,dashed](-.8,0)--(.3,0)node[black,left=.8cm,above=.2cm]{$H$};
\draw[very thick](.5,0)--(3,0)node[black,left=2cm,above=.2cm]{$(0,1)$};
\draw[fill=red] (-.9,0) circle [radius=.1] node[black,font=\large,below=.1cm]{$\bar{a}_1$};
\draw[fill=red] (.4,0) circle [radius=.1] node[black,font=\large,below=.1cm]{$\bar{a}_2$};
\end{tikzpicture}}
\hspace{2em}
\subfloat[{ Black ring spacetime}]{
\begin{tikzpicture}[scale=.7, every node/.style={scale=0.7}]
\draw[very thick](-4,0)--(-.9,0)node[black,left=2cm,above=.2cm]{$(1,0)$};
\draw[thick,dashed](-.7,0)--(.7,0)node[black,left=.8cm,above=.2cm]{$H$};
\draw[very thick](.9,0)--(1.7,0)node[black,left=.3cm,above=.2cm]{$(1,0)$};
\draw[very thick](1.9,0)--(4,0)node[black,left=1cm,above=.2cm]{$(0,1)$};
\draw[fill=red] (-.8,0) circle [radius=.1] node[black,font=\large,below=.1cm]{$\bar{a}_1$};
\draw[fill=red] (.8,0) circle [radius=.1] node[black,font=\large,below=.1cm]{$\bar{a}_2$};
\draw[fill=red] (1.8,0) circle [radius=.1] node[black,font=\large,below=.1cm]{$\bar{a}_3$};
\end{tikzpicture}}
\vspace{1em}
\subfloat[{Orbit space of a Myers-Perry spatial slice on upper half plane}]{
\begin{tikzpicture}[scale=.8, every node/.style={scale=0.7}]
\fill[ black!50!blue, opacity=0.35](-4,0)--(4,0)--(4,3)--(-4,3);
\draw[very thick](0,0)--(0,3)node[black,above=.2cm]{$\rho$};
\draw[very thick](-4,0)--(-.1,0)node[black,left=2cm,above=.2cm]{$(1,0)$}node[black,left=2cm,below=.2cm]{$I_-$};
\draw[very thick](.1,0)--(4,0)node[black,left=2cm,above=.2cm]{$(0,1)$}node[black,left=2cm,below=.2cm]{$I_+$}node[black,right=.2cm]{$z$};
\draw[fill=white] (0,0) circle [radius=.1] node[black,font=\large,below=.1cm]{${a}_E$};
\end{tikzpicture}}
\hspace{2em}
\subfloat[{Orbit space of nonextereme black ring spatial slice as infinite strip }]{
\begin{tikzpicture}[scale=.8, every node/.style={scale=0.7}]
\fill[ black!50!blue, opacity=0.35](-4,0)--(4,0)--(4,3)--(-4,3);
\draw[very thick](0,0)--(0,3)node[black,above=.2cm]{$\rho$};
\draw[very thick](-4,0)--(-.1,0)node[black,left=2cm,above=.2cm]{$(1,0)$};
\draw[very thick](.1,0)--(4,0)node[black,left=4cm,above=.2cm]{$(0,1)$}node[black,left=2.6cm,above=.2cm]{$(1,0)$}node[black,left=.6cm,above=.2cm]{$(0,1)$}node[black,right=.2cm]{$z$};;
\draw[fill=white] (0,0) circle [radius=.1] node[black,font=\large,below=.1cm]{${a}_E$};
\draw[fill=red] (1,0) circle [radius=.1] node[black,font=\large,below=.1cm]{${a}_2$};
\draw[fill=red] (2.3,0) circle [radius=.1] node[black,font=\large,below=.1cm]{${a}_1$};
\end{tikzpicture}}
\caption{(a) and (b) are spacetime interval structures for the Myers-Perry black hole and Emparan-Reall black ring. (c) and (d) depict the orbit spaces  for the corresponding complete initial data maximal slices with a spherical minimal surface and a ring-topology minimal surface respectively. The rod point ${a}_E$ in axis represents another asymptotic end of the slice.}\label{fig1}
\end{figure}
If we are considering initial data for spacetimes containing black holes, then as discussed above, $\Sigma$ will have additional asymptotic ends. Hence the orbit space will have, in addition to a boundary consisting of intervals and an asymptotic boundary, additional points removed from it. By symmetry, these points must lie on the axis $\rho =0$. Of course, these removed points represent entire asymptotic regions that are an infinite proper distance from other points in $\Sigma$. Note that a similar situation in the Lorentzian setting occurs when analyzing the orbit space of extreme black holes, when the rod corresponding to the timelike Killing field becoming null shrinks to zero size \cite{Figueras2010}. In summary,  the boundary $\rho =0$ of the orbit space for more general initial data will consist of two semi-infinite rods, possibly finite rods, and points removed between the rods.  We illustrate this in Figure \ref{fig1}, which shows the orbit space of a black hole spacetime and its associated standard constant time maximal slice for the Myers-Perry black hole\cite{myers1986black} and a non-extreme 
black ring \cite{emparan2006black}.  

\begin{figure}
\centering
\subfloat[{Orbit space of a Myers-Perry spatial slice represented on infinite strip }]{
\begin{tikzpicture}[scale=.85, every node/.style={scale=0.6}]
\draw[black,thick](-3,0)--(3.3,0)node[black,font=\large,right=.2cm]{$y$};
\draw[black,thick](0,-2)--(0,2)node[black,font=\large,above=.2cm]{$x$};
\fill[ black!50!blue, opacity=0.35](-3,1)--(3,1)--(3,-1)--(-3,-1);
\draw[black,ultra thick](-3,1)--(3,1)node[black,font=\large,above=.5cm,left=2cm]{$(0,1)$}node[black,font=\large,right=.2cm]{$\mathcal{I}^+$};
\draw[black,ultra thick](-3,-1)--(3,-1)node[black,font=\large,below=.5cm,left=2cm]{$(1,0)$}node[black,font=\large,right=.2cm]{$\mathcal{I}^-$};
\end{tikzpicture}}
\\
\subfloat[{Orbit space of a non-extreme black ring slice on the infinite strip }]{
\begin{tikzpicture}[scale=.85, every node/.style={scale=0.6}]
\draw[black,thick](-3,0)--(3.3,0)node[black,font=\large,right=.2cm]{$y$};
\draw[black,thick](0,-2)--(0,2)node[black,font=\large,above=.2cm]{$x$};
\fill[ black!50!blue, opacity=0.35](-3,1)--(3,1)--(3,-1)--(-3,-1);
\draw[black,ultra thick](-3,1)--(3,1)node[black,font=\large,above=.5cm,left=1cm]{$(0,1)$}node[black,font=\large,right=.2cm]{$\mathcal{I}^+$}node[black,font=\large,above=.5cm,left=4.3cm]{$(1,0)$}node[black,font=\large,above=.5cm,left=6.8cm]{$(0,1)$};
\draw[black,ultra thick](-3,-1)--(3,-1)node[black,font=\large,right=.2cm]{$\mathcal{I}^-$}node[black,font=\large,below=.5cm,left=2cm]{$(1,0)$};
\draw[red,fill=red] (-1.5,1) circle [radius=.07] node[black,above=.1cm]{$a_{2}$};
\draw[red,fill=red] (.7,1) circle [radius=.07] node[black,above=.1cm]{$a_{1}$};
\end{tikzpicture}}
\hspace{3em}
\subfloat[{Orbit space of an extreme black ring slice on infinite strip }]{
\begin{tikzpicture}[scale=.85, every node/.style={scale=0.6}]
\draw[black,thick](-3,0)--(3.3,0)node[black,font=\large,right=.2cm]{$y$};
\draw[black,thick](0,-2)--(0,2)node[black,font=\large,above=.2cm]{$x$};
\fill[ black!50!blue, opacity=0.35](-3,1)--(3,1)--(3,-1)--(-3,-1);
\draw[black,ultra thick](-3,1)--(3,1)node[black,font=\large,above=.5cm,left=2cm]{$(0,1)$}node[black,font=\large,right=.2cm]{$\mathcal{I}^+$}node[black,font=\large,above=.5cm,left=6cm]{$(1,0)$};
\draw[black,ultra thick](-3,-1)--(3,-1)node[black,font=\large,right=.2cm]{$\mathcal{I}^-$}node[black,font=\large,below=.5cm,left=2cm]{$(1,0)$};
\draw[red,fill=red] (-.7,1) circle [radius=.07] node[black,above=.1cm]{$a_{1}$};
\end{tikzpicture}}
\caption{The doubling of extreme slice yield to non-extreme slice with double orbit space. Here $y=\log r$} \label{fig2}
\end{figure}
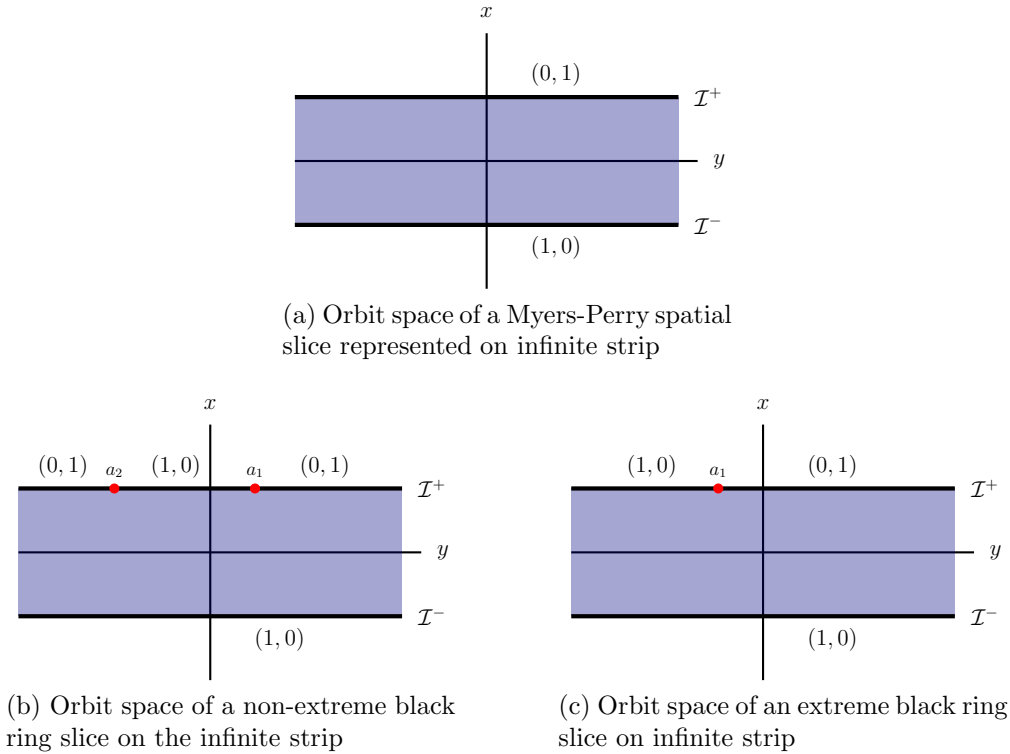

In order to understand the topology of these asymptotic ends (e.g. to distinguish between spatial slices of the non-extreme and extreme black rings) it proves useful to conformally map the upper half plane ($\rho \geq 0, -\infty < z < \infty$) to an infinite strip parameterized by $0 < r < \infty, -1 \leq x \leq 1$ via the transformation
\begin{equation}\label{rx}
\rho = \frac{r^2}{2} \sqrt{1-x^2} \; ,  \qquad z = \frac{r^2}{2} x \; . 
\end{equation}  There are now between two to four (possibly semi)-infinite intervals, corresponding to \emph{two} asymptotic ends at $r=0$ and as $r \to \infty$.   The asymptotically flat end at large $r$ must have different Killing vector fields vanishing on the upper $x=1$ and lower ($x=-1$) axis, corresponding to an asymptotic $S^3$ boundary. However, at the end $r=0$ we could have either an asymptotic $S^3$ (corresponding to another asymptotically flat end, or an asymptotically cylindrical end with $N \cong S^3$) or an asymptotic $S^1 \times S^2$ (corresponding to an asymptotically cylindrical end with ring topology, as arises for a complete spatial slice of an extreme black ring).  More generally one may obtain an $N\cong L(p,q)$. Further details will be discussed in the following sections.  A schematic illustration of the infinite strip representation of the orbit space for spatial slices of the Myers-Perry black hole and  black rings are given in Figure \ref{fig2}

\par We now discuss the behaviour of the functions appearing in the class of conformal metrics \eqref{idmet} on the boundary and asymptotic regions of the orbit space.  These will be required to analyze properties of the mass functional to be defined in the next section.

\subsubsection{Behaviour at the asymptotically flat end}
First of all, note that the Euclidean metric on $\mathbb{R}^4$ in the $\rho,z$ chart given in \eqref{idmet} is
\begin{equation}\label{flatmet}
\delta_4 = \frac{\td\rho^2 + \td z^2}{2\sqrt{\rho^2 + z^2} } + (\sqrt{\rho^2 + z^2} - z) \td\phi^2 + (\sqrt{\rho^2 + z^2} + z) \td\psi^2
\end{equation} where $\rho \in \mathbb{R}^+\cup\{0\}$ and $z \in \mathbb{R}$. This can be put in a more familiar chart by using the transformation \eqref{rx} and noting that $r^2 = 2 \sqrt{\rho^2 + z^2}$, the metric is
\begin{equation}\label{flatmetric}
\delta_4 = \td r^2 +  \frac{r^2 \td x^2}{4(1-x^2)} + \frac{r^2}{2} ((1-x) \td\phi^2 + (1+x) \td\psi)
\end{equation} where $r\geq 0$ and $-1 \leq x \leq 1$ and  $\phi,\psi$ have period $2\pi$.  Hence our asymptotically flat metrics must approach $\delta_4$ with appropriate fall-off conditions. Note that asymptotic infinity corresponds to $r \to \infty$ so that $\rho, z \to \infty$ with $z (\rho^2 + z^2)^{-1/2}$ fixed and the axes of rotation $x= \pm 1$ lie on the axis $\rho = 0$ with finite $z$ \cite{Figueras2010}. In particular, the boundary of the orbit space for $(\mathbb{R}^4,\delta_4)$ consists of the semi-infinite rods $I_-: -\infty < z < 0$ and $I_+: 0 < z < \infty$  where $\partial_\psi$ and $\partial_\phi$  vanish respectively.

Let us now consider our class of asymptotically flat  conformal metrics $\tilde{h}$.  We will consider $U$ and $\lambda'_{ij}$ as functions on the orbit space $\mathcal{B}$.  First of all, asymptotic flatness implies $e^{-2U} \to 2 \sqrt{\rho^2 + z^2}$. Since we assume the conformal metric has zero ADM mass, it is convenient to decompose $U$ as
\begin{equation}
U = V - \frac{1}{2} \log\left(2 \sqrt{\rho^2 + z^2}\right)\label{confU} 
\end{equation} where $V= O(r^{-2})$, that is
\begin{equation}
V = \frac{\bar{V}(x)}{r^2} + o(r^{-2}) \; ,  \qquad r \to \infty\label{Vinfty}
\end{equation}  and $\bar{V}$ satisfies the condition that the integral given in \eqref{Vbar} vanishes. As shown in Appendix \ref{appB}, this is equivalent to the requirement that $\tilde{h}$ has vanishing ADM mass. 
Next, we take the fall-off conditions of the Killing metric $\lambda'_{ij}$ to be
\begin{equation}\label{lambdainfty}
\lambda'_{11}=\frac{r^2}{2}(1-x)[1+ \frac{f(x)}{r^2} + o(r^{-2})]\, , \quad \lambda'_{22}=\frac{r^2}{2}(1+x)[1+\frac{g(x)}{r^2} + o(r^{-2})]\, ,\quad \lambda'_{12}=(1-x^2)o(r^{-2})
\end{equation}  with $f(x)+ g(x)= 0$ because $\det \lambda' = \rho^2$ where $\rho$ is given by \eqref{rx}.   We also assume the following fall off at infinity $r\to\infty$
\begin{equation}
Y^1 = y_1-\frac{J_1(x+1)^2}{\pi}+ \mathcal{O}(r^{-2}) \quad Y^2 = y_2-\frac{J_2(3-x)(x+1)}{\pi}+ \mathcal{O}(r^{-2})\label{Yinfty}
\end{equation}
where $J_i$ are angular momenta and $y_i$ are constants \cite{Figueras2010}.  Therefore we have
\begin{gather}
\tilde{K}_{ab}= o(1/r^3)\label{Kinf} \;. 
\end{gather}  
Finally, we have assumed
\begin{equation}
\Phi-1=\mathcal{O}(1/r^2)\qquad \Phi_{,r}=\mathcal{O}(1/r^3)\label{coninf}
\end{equation} which is sufficient to ensure finite ADM mass of $(\Sigma,h)$.
\subsubsection{Boundary conditions on the axis}
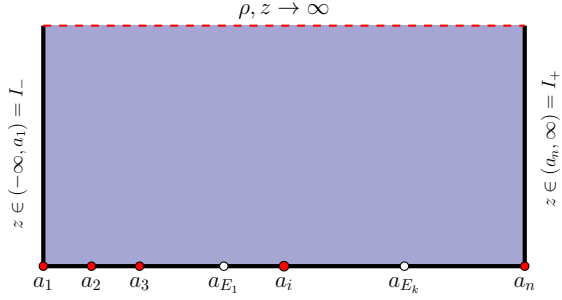
\begin{figure} 
\centering
\begin{tikzpicture}[scale=.8, every node/.style={scale=0.6}]
\draw[ fill=black!50!blue, opacity=0.35,very thick](-4,4)--(-4,0)node[rotate=90,black,opacity=1,right=2.5cm,above=.2cm]{$z\in(-\infty,a_1)=I_-$}--(4,0)--(4,4)node[rotate=90,black,opacity=1,left=2.5cm,below=.2cm]{$z\in(a_n,\infty)=I_+$};
\draw[ultra thick, black](-4,4)--(-4,0)--(4,0)--(4,4);
\draw[thick,red, dashed](-4,4)--(4,4)node[black,midway,font=\large,above]{$\rho,z\to\infty$};
\draw[fill=red] (-4,0) circle [radius=.07] node[black,font=\large,below=.1cm]{$a_1$};
\draw[fill=red] (-3.2,0) circle [radius=.07] node[black,font=\large,below=.1cm]{$a_2$};
\draw[fill=red] (-2.4,0) circle [radius=.07] node[black,font=\large,below=.1cm]{$a_3$};
\draw[fill=white] (-1,0) circle [radius=.07] node[black,font=\large,below=.1cm]{$a_{E_1}$};
\draw[fill=red] (0,0) circle [radius=.08] node[black,font=\large,below=.1cm]{$a_{i}$};
\draw[fill=white] (2,0) circle [radius=.07] node[black,font=\large,below=.1cm]{$a_{E_k}$};
\draw[fill=red] (4,0) circle [radius=.07] node[black,font=\large,below=.1cm]{$a_n$};
\end{tikzpicture}
\caption{ The {\color{blue} blue} region is orbit space $\mathcal{B}$. The {\bf black} line is axis $\rho=0$ and {\color{red} red} dashed line is infinity $z,\rho \to \infty$. The rod points $a_{E_k}$ represent other asymptotic ends.} \label{fig3}
\end{figure}

The  boundary of the orbit space $\partial \mathcal{B}$ lies on the z-axis  $\rho=0$.
 We know by \cite{harmark2004stationary} that the eigenspace for the eigenvalue zero of the matrix $\lambda'_{ij}$ for a given $z$ is one-dimensional, except for isolated values of $z$. These isolated points  are denoted $a_1,\cdots,a_n$ and we can divide the axis into subintervals $(-\infty,a_1),(a_1,a_2),\cdots,(a_n,\infty)$. On each interval a particular integer linear combination of the $m_i$ vanishes.  The semi-infinite rods $I_\pm$ at the ends correspond to the axes of rotation of the asymptotic $\mathbb{R}^4$ region.  Without loss of generality, we can choose $m_1, m_2$ to vanish on $I_+$ and $I_-$ respectively. A sketch of a typical orbit space is given in Figure \ref{fig3}.
 
 The finite-length rods, on the other hand,  correspond to 2-cycles in $\Sigma$.  Suppose that on a particular rod $I_i$ we have
 \begin{equation}
\lambda'_{ij}v^j=0 \; , \qquad v=v^i\frac{\partial}{\partial\phi^i} \ , \qquad v^i \in \mathbb{Z}
\end{equation} By an $SL(2,\mathbb{Z})$ change of basis $(m_1,m_2) \to (m'_1, m'_2)$ of the Killing fields, one can always choose $v = m'_1$ and another basis vector $w = m'_2$ such that $w$ is non-vanishing on $I_i$ except at its two endpoints $a_i$ and $a_{i+1}$. Hence the interval $I_i$ is topologically an $S^2$ submanifold of $\Sigma$.


The functions $V$ and $\lambda'_{ij}$ must satisfy regularity requirements as $\rho \to 0$. We will review these briefly here.  
  Let $\psi$ be a coordinate such that $\partial/ \partial \psi=v$. Then in order to ensure the absence of  conical singularities of the metric we impose that 
\begin{equation}
\Delta \psi =2\pi\lim_{\rho\to 0}\sqrt{\frac{\rho^2e^{2U}}{\lambda'_{ij}v^iv^j}}=2\pi\lim_{\rho\to 0}\sqrt{\frac{\rho^2e^{2V}}{2\sqrt{\rho^2+z^2}\lambda'_{ij}v^iv^j}}\nonumber\\
=2\pi\qquad z\in(a_i,a_{i+1})
\end{equation} and hence
\begin{eqnarray}\label{regaxis}
V=\frac{1}{2}\log\left(\frac{2\sqrt{\rho^2+z^2}\lambda'_{ij}v^iv^j}{\rho^2}\right)=\frac{1}{2}\log V_i\quad\quad\text{where}\quad z\in(a_i,a_{i+1})\mbox{ and } \rho\to 0\label{originV}
\end{eqnarray}
where $V_i=V_i(z)$ is a bounded function.

Now consider the behaviour of $V$ on the semi-infinite rods $I_\pm$ defined above. The metric must take the asymptotic form of the flat metric. Then on either axis, we have, as $z \to \pm \infty$,
\begin{equation}
\lambda'_{ij}v^i v^j = \frac{1}{2|z|} \rho^2 + O(\rho^4) \, \label{lambdaaxis}
\end{equation}  where $v$ is taken to be $m_1$ or $m_2$. Thus we see that $V_\pm  \to 1$ as $z \to \pm \infty$, where  $V_\pm$ refers to the function $V_i$ evaluated on $I_\pm$.  

Moreover,  consider the rod $I_i$ (either finite or semi-infinite). Near $I_i$, in the adapted basis $(m'_1,m'_2)$  discussed above we have the following behaviour for $\lambda'_{ij}$
\begin{equation} \label{lambdaaxis}
\lambda'_{ij} =
 \begin{pmatrix}
  \mathcal{O}(\rho^2) & \mathcal{O}(\rho^2) \\
 \mathcal{O}(\rho^2) & \mathcal{O}(1)
 \end{pmatrix}\qquad \qquad\rho \to 0
\end{equation}
In addition, we will require that near $I_i$,  the twist potentials in this basis behave as 
\begin{equation}\label{twistaxis}
Y^1=C_1+\mathcal{O}(\rho^4)\qquad  Y^2=C_2+\mathcal{O}(\rho^2)\qquad \qquad\rho \to 0
\end{equation}
where $C_1$ and $C_2$ are constants.  The fall-off of $Y^1$ is more restrictive than simply requiring $\td Y^1 = \mathcal{O}(\rho^2)$ along a rod where $v = m'_1$ vanishes, but this condition is satisfied by the twist potentials of the Myers-Perry and black ring solutions.  The condition \eqref{twistaxis} will be needed to show certain terms in the mass functional are finite as $\rho \to 0$. 

Finally, on the axis $\rho = 0$, apart from isolated points corresponding to asymptotic ends, we require
\begin{equation}
\Phi=\mathcal{O}(1) \; . 
\end{equation}
\subsubsection{Behaviour near asymptotic ends}
As discussed above, $\Sigma$ may have additional asymptotic ends. Note that to have non -trivial angular momenta, $\Sigma$ must have non trivial topology.  More precisely, if $S$ represents the sphere at infinity, then since $\td \star \td m_i =0$ by virtue of the vacuum spacetime equations, it follows that the Komar angular momenta $J_i$ vanishes unless $S$ is \emph{not} the boundary of some compact domain contained in $\Sigma$.  Such a situation arises when isolated points are removed from $\Sigma$, yielding additional asymptotic ends.  By the $U(1)^2$ symmetry assumption, these lie on a point on the axis $\rho=0$.  For example, in the case that the initial data arises from a stationary black hole,  the location of the removed point corresponds to the location of the event horizon.  We will allow for both asymptotically flat ends and cylindrical ends, which the latter arise in the context of initial data for extreme black holes \cite{alaee2014small}. We impose singular boundary conditions on the conformal factor $\Phi$:
\begin{eqnarray}\label{conformalend}
\Phi &=& \mathcal{O}(r_i^{-2}) \qquad \partial_{r_i}\Phi = \mathcal{O}(r_i^{-3}) \quad \mbox{asymptotically flat} \\
\Phi &=& \mathcal{O}(r_i^{-1}) \qquad \partial_{r_i}\Phi = \mathcal{O}(r_i^{-2}) \quad \mbox{cylindrical end}
\end{eqnarray} where $r_i$ represents the distance to the asymptotic end.  We assume that the conformal metric $\tilde{h}$  approaches the flat metric on $\mathbb{R}^4$ in the former case, whereas in the latter case, the conformal metric $\tilde{h}$ will be asymptotically cylindrical, so $\tilde{h}= \Omega^2 ( \td r_i^2 + r_i^2 \gamma)$ where $\gamma$ is a metric on a compact manifold (the example of extreme Myers-Perry is discussed in detail in Appendix B).  This assumption is most easily illustrated by considering the example of a maximal constant-$t$ slice of the five-dimensional Schwarzschild geometry,
\begin{equation}
h = \left( 1 + \frac{\mu}{2r^2}\right)^2 \delta_4
\end{equation} where $\tilde{h} = \delta_4$ is the flat metric on $\mathbb{R}^4$ given in \eqref{flatmet} and $r= 2 \sqrt{\rho^2 + z^2}$.  One easily sees that the point $r=0$ corresponds to another asymptotically flat end and the conformal factor $\Phi$ has the singular behaviour \eqref{conformalend}. In this simple case, this asymptotic region corresponds to the corner point $(\rho,z) = (0,0)$ on the axis of the orbit space where both Killing fields vanish. 

For the initial data of the Myers-Perry black hole, the conformal factor $\Phi$ diverges in the same way and the removed point is again located at a corner  \cite{hollands2008uniqueness} of the orbit space (the point $\bar{a}_E$ in Figure \ref{fig1} (c)).  However, for the black ring, the removed point is \emph{not} at a corner, but instead lies on the rod corresponding to the Killing field which vanishes on the $S^2$ of the horizon. One can again verify that $\Phi$ has the above singular behaviour at this point, and is in fact regular at the point $\bar{a}_2$ (see Figure \ref{fig1} (d)).
\section{A mass functional for initial data}\label{masssec}
We now follow the approach of Dain \cite{dain2006variational} to construct a proposal for a mass functional $\mathcal{M}$. This functional  depends on the functions $(\lambda'_{ij}, Y^i)$ and should evaluate to the mass for the class of initial data we considered in the previous section when $t-\phi^i$ symmetry holds.  Our starting point is the ADM mass for the asymptotically flat, complete  Riemannian manifold $(\Sigma,h_{ab})$:
\begin{equation}\label{ADMmass}
M_{ADM}=\frac{1}{16\pi}\lim_{r\rightarrow\infty}\int_{S^3_r}\left(\partial_ah_{ac}-\partial_ch_{aa}\right)n^c\, ds_h
\end{equation} where $S^3_r$ refers to a three-sphere of coordinate radius $r$ with volume element $ds_h$ in the Euclidean chart outside a large compact region and $n$ is the unit normal.  If we evaluate the ADM mass in terms of the conformally scaled initial data and by the assumptions in the previous section , one has
\begin{equation}
 M_{ADM}=-\frac{3}{8\pi}\lim_{r\rightarrow\infty}\int_{S^3_r}n^c \tilde{\nabla}_c\Phi \, ds_{\tilde{h}}\label{conformalmass1}
\end{equation} where $\tilde{\nabla}$ refers to the covariant derivative with respect to $\tilde{h}_{ab}$.  Using \eqref{eq:constraints} and the fact that $\Phi \to 1$ as $r \to \infty$,  define
\begin{eqnarray}
m
&\equiv& -\frac{3}{8\pi}\lim_{r\rightarrow\infty}\int_{S^3_r}\frac{\tilde{\nabla}_c\Phi}{\Phi}\, n^c\, ds_{\tilde{h}}  \nonumber\\
&=& -\frac{3}{8\pi}\int_{\Sigma}\tilde{\nabla}^c\left(\frac{\tilde{\nabla}_c\Phi}{\Phi}\right)\, \, d\Sigma_{\tilde{h}}+\frac{3}{8\pi}\lim_{r_i\rightarrow 0}\int_{N_i}\frac{\tilde{\nabla}_c\Phi}{\Phi}\, n^c\, ds_{\tilde{h}} \nonumber\\
&=& \frac{3}{8\pi}\int_{\Sigma}\left(-\frac{\tilde{R}}{6}+\frac{\tilde{K}_{ab}\tilde{K}^{ab}}{6\Phi^{6}}+\frac{\tilde{\nabla}^c\Phi\tilde{\nabla}_c\Phi}{\Phi^2}\right)\, \, d\Sigma_{\tilde{h}} +\frac{3}{8\pi}\lim_{r_i\rightarrow 0}\int_{N_i}\frac{\tilde{\nabla}_c\Phi}{\Phi}\, n^c\, ds_{\tilde{h}}
\end{eqnarray} where in passing from the first line to the second line, we have used the divergence theorem.  Provided the behaviour of $\Phi$ at the asymptotic ends is given by   \eqref{conformalend} the last term in $m$ is zero.  Here $N_i$ refers to the geometry of the $i$th asymptotic end (this can be $S^3, S^1 \times S^2$ or more generally $L(p,q)$).

This form of expressing the mass as a bulk integral is important for defining the functional $\mathcal{M}$.  For our class of initial data, we can reduce the integral to one over  the orbit space $\mathcal{B}$.  Note that $\det\tilde{h} =  e^{4U}\rho^2$. Performing the trivial integrals over the angles gives
\begin{eqnarray}\label{m}
m  &=&\frac{3\pi}{2}\int_{\mathcal{B}}\left(-\frac{\tilde{R}e^{2U}}{6}+\frac{\tilde{K}_{ab}\tilde{K}^{ab}e^{2U}}{6\Phi^{6}}+\frac{\partial_A\Phi\partial^A\Phi}{\Phi^2}\right)\, \, \rho\td\rho\td z
\end{eqnarray} Equivalently,  in terms of the flat metric on $\mathbb{R}^3$ in cylindrical coordinates
\begin{equation}\label{3flat}
\delta = \td \rho^2 + \td z ^2 + \rho^2 \td\varphi^2
\end{equation} where $\varphi$ is an auxiliary angular coordinate with period $2\pi$, we can write 
\begin{eqnarray}
m  &=&\frac{3}{4}\int_{\mathbb{R}^3}\left(-\frac{\tilde{R}e^{2U}}{6}+\frac{\tilde{K}_{ab}\tilde{K}^{ab}e^{2U}}{6\Phi^{6}}+\frac{\partial_A\Phi\partial^A\Phi}{\Phi^2}\right)\, \, \td\mu_0 \nonumber
\end{eqnarray} and $\td\mu_0$ is the volume element of $\delta$. 

We are now in the position to define our mass functional for arbitrary $t-\phi^i$ symmetric initial data. Recall for this data, the extrinsic curvature is specified in terms of twist potentials $Y^i$ as given by \eqref{exttphi} and its square is given by the contraction \eqref{tildeK^2}. Using the expression for the scalar curvature  $\tilde{R}$  \eqref{scalarcurvature} we have
\begin{eqnarray}\label{mass3d}
m=\frac{1}{8}\int_{\mathbb{R}^3}\left(2\Delta_2 U-\frac{1}{\rho^2}+\frac{1}{4}\text{Tr}\left[\left(\lambda'^{-1}\td \lambda'\right)^2\right]+e^{-6v}\frac{\text{Tr}\left(\lambda'^{-1}\td Y\td Y^t\right)}{2\det\lambda'}+6\left(\td v\right)^2\right)\, \, \td\mu_0
\end{eqnarray}  where $v=\log\Phi$. As an integral over $\mathbb{R}^3$, this expression appears similar to the analogous formula for $m$ when $N=3$  first given in \cite{dain2006variational}.  However, in $N=4$ there are a number of key qualitative differences. 

Firstly, consider the terms
\begin{equation}
-\frac{1}{\rho^2}+\frac{1}{4}\text{Tr}\left[\left(\lambda'^{-1}\td \lambda'\right)^2\right] \; .
\end{equation} In $N=3$, it is easily seen that the above expression vanishes identically. This is no longer the case in $N=4$. We note for later use the identity: 
\begin{eqnarray}
-\frac{1}{\rho^2}+\frac{1}{4}\text{Tr}\left[\left(\lambda'^{-1}\td \lambda'\right)^2\right]
&=&-\frac{1}{4}\left(\text{Tr}\left(\lambda'^{-1}\td \lambda'\right)\right)^2+\frac{1}{4}\text{Tr}\left[\left(\lambda'^{-1}\td \lambda'\right)^2\right]\nonumber\\
&=&-\frac{1}{2}\frac{\det\td \lambda'}{\det\lambda'}\label{2Didentity} \qquad\text{for $2 \times 2$ matrices}
\end{eqnarray} where we are using the notation $\det \td\lambda' = \tfrac{1}{2}\epsilon^{ik} \epsilon^{jl} \td\lambda'_{ij} \cdot \td\lambda'_{kl} $.

A second important difference is that, unlike in $N=3$, the integral over $\Delta_2 U$ does \emph{not} vanish.  Indeed,  in terms of the rod structure formalism, the class of three-dimensional initial data studied in \cite{dain2006variational} has $\partial \mathcal{B}$  consisting of a single rod (the rotation axis of the generator of the $U(1)$ symmetry) with points removed corresponding to asymptotic ends.  As we shall now explain, this is sufficient, along with appropriate fall-off conditions, to prove that the first term in $m$ does not contribute either.  However, we shall see the situation is more complicated in our present case. Note that
\begin{equation}
\Delta_2 U = \frac{\partial^2 U}{\partial \rho^2} + \frac{\partial^2 U}{\partial z^2}
\end{equation} 
Using our expression for $U$, we have
\begin{eqnarray}
\int_{\mathcal{B}}\Delta_2U\rho\,\td\rho\td z&=&\int_{\mathcal{B}}\left(\Delta_2V-\frac{1}{2}\Delta_2\log\left(2\sqrt{\rho^2+z^2}\right)\right)\rho\,\td\rho\td z\nonumber\\
&=&\int_{\mathcal{B}}\Delta_2V\rho\,\td\rho\td z = \int_{\mathcal{B}}\td\alpha 
\end{eqnarray}
where we have defined the one-form
\begin{equation} \label{alpha}
\alpha \equiv  \left(\rho V_{,\rho}-V\right)\td z-\rho V_{,z}\td \rho
\end{equation} Recall the boundary of the orbit space consists of the asymptotic region 
$\mathcal{B}_\infty \equiv \{z,\rho \to \infty, z (\rho^2 + z^2)^{-1/2} \textrm{ finite} \}$, i.e. $r\to\infty$, and the axis $\rho =0$ denoted by $\partial\mathcal{B}$.   Using the asymptotic condition \eqref{Vinfty}, we find to leading order $\alpha = -\bar{V}(x) \td x$ as $r \to \infty$.  Hence by Stokes' theorem we have
\begin{eqnarray}
\int_{\mathcal{B}} \td \alpha &=&\int_{\partial\mathcal{B} \cup \mathcal{B}_\infty }\alpha =\int_{\partial\mathcal{B} }\alpha = \int_{I_- \cup I_1 \cup\cdots\cup I_{+}} \alpha \nonumber \\
&=&\int_{I_- \cup I_1 \cup\cdots\cup I_{+}}\left(\rho V_{,\rho}-V\right)|_{\rho=0}\,\td z =- \int_{I_- \cup I_1 \cup\cdots\cup I_{+}}V|_{\rho=0}\,\td z\nonumber\\
&=&-\frac{1}{2}\sum_{\text{rods}}\int_{I_i}\log V_i\,\td z\nonumber
\end{eqnarray}  Note that the integral over $\mathcal{B}_\infty$ vanishes as a consequence of the condition $\tilde{h}$ has vanishing ADM mass \eqref{Vbar}. 

The above considerations lead us to define the following mass functional for $t-\phi^i$-symmetric, maximal asymptotically flat vacuum data:
\begin{eqnarray}
\mathcal{M}
&\equiv&\frac{\pi}{4}\int_{\mathcal{B}}\left(-\frac{1}{\rho^2}+\frac{1}{4}\text{Tr}\left[\left(\lambda'^{-1}\td \lambda'\right)^2\right]+e^{-6v}\frac{\text{Tr}\left(\lambda'^{-1}\td Y\td Y^t\right)}{2\det\lambda'}+6\left(\td v\right)^2\right)\, \,\rho \td\rho \td z \nonumber \\ &-&\frac{\pi}{4}\sum_{\text{rods}}\int_{I_i}\log V_i\,\td z\label{mass}
\end{eqnarray}  where $\mathcal{M} = \mathcal{M}(\lambda'_{ij}, Y^i, v) $.  Note that if we consider maximal, $U(1)^2$-invariant data without $t-\phi^i$ symmetry, we have $m \geq \mathcal{M}$ as a consequence of \eqref{Kaxi}. 

The mass functional depends on the matrix $\lambda'_{ij}$, the twist potentials $Y^i$, and the conformal factor $v$. It also depends upon the \emph{boundary} values of the  $\lambda'_{ij}$ along finite rods on the axis, via the functions $V_i$. Define $\mathcal{A}\equiv\{(\lambda', Y,v ): \text{$\mathcal{M}(\lambda', Y,v )$ is bounded}\}$, then $\mathcal{M}$ will be well-defined on $\mathcal{A}$.   Of course, not all elements belonging to $\mathcal{A}$ will represent the mass of some initial data set. There are three functions in $\lambda'_{ij}$ with a constraint $\det\lambda'=\rho^2$ so there are two independent functions in $\lambda'_{ij}$,
 two independent potentials $Y^i$ and conformal factor $\Phi$ (or $v=\log\Phi$). We have seen that all axisymmetric and $t-\phi^i$ symmetric data can be
generated by six functions $(U,\lambda', Y,v )$. These functions are coupled by the Lichnerowiscz equation \eqref{eq:Lich} which can be rewritten as 
\begin{equation}
\Delta_{3}v+\frac{1}{3}\Delta_2U-\frac{1}{6\rho^2}+\frac{1}{24}\text{Tr}\left[\left(\lambda'^{-1}\td \lambda'\right)^2\right]=e^{-6v}\frac{\text{Tr}\left(\lambda'^{-1}\td Y\td Y^t\right)}{12\det\lambda'}
\end{equation}
where $\Delta_{3}$ is three dimensional Laplace operator with respect to metric $\delta$.  Now for given $(\lambda'_{ij},Y, v)$, we have a linear two dimensional Poisson equation for $U$. Then by equation \eqref{confU} we have a linear two dimensional Poisson equation for $V$:
\begin{gather}
\Delta_2V=F(v,\lambda',Y)\label{25}
\end{gather} 
with boundary  \eqref{Vinfty} at infinity and we have $V=O(1)$ at $\rho=0$. Now let $\mathcal{A}_1$ be a solution of equation \eqref{25} with appropriate fall- off conditions. Then $\mathcal{M}(\lambda'_{ij}, Y^i, v)$  will give us the mass of a initial data set only if the given data is selected from $\mathcal{A}_1 \subset \mathcal{A}$.  


Finally,  using the asymptotic and boundary conditions on the orbit space functions, we now show that $\mathcal{M}$ is finite. By asymptotic condition \eqref{lambdainfty} the behaviour of the first two terms of $\mathcal{M}$ near infinity is
\begin{eqnarray}
-\frac{1}{\rho^2}+\frac{1}{4}\text{Tr}\left[\left(\lambda'^{-1}\td \lambda'\right)^2\right]
=\mathcal{O}(r^{-8})\qquad\text{as}\qquad r\to\infty
\end{eqnarray}
Thus it is bounded at infinity.  Near the axis, we must analyze the behaviour of these terms near each rod. One can check
\begin{eqnarray}
-\frac{1}{\rho^2}+\frac{1}{4}\text{Tr}\left[\left(\lambda'^{-1}\td \lambda'\right)^2\right]
=\mathcal{O}(1)\qquad\text{as}\qquad \rho\to 0
\end{eqnarray}  The third term in the mass functional by equations \eqref{lambdainfty} and \eqref{Yinfty}  has following behaviour at infinity
\begin{eqnarray}
e^{-6v}\frac{\text{Tr}\left(\lambda'^{-1}\td Y\td Y^t\right)}{2\det\lambda'}=\mathcal{O}(r^{-10})\qquad\text{as}\qquad r\to\infty
\end{eqnarray} and near the axis one has, using \eqref{lambdaaxis} and \eqref{twistaxis},
\begin{eqnarray}
e^{-6v}\frac{\text{Tr}\left(\lambda'^{-1}\td Y\td Y^t\right)}{2\det\lambda'}=\mathcal{O}(1)\qquad\text{as}\qquad \rho\to 0
\end{eqnarray} Finally, if one uses the conditions near additional asymptotically flat ends, one can  ensure that $\mathcal{M}$ is finite, assuming continuity of the functions in the interior of $\mathcal{B}$. 
\section{Stationary, biaxisymmetric data}
Let us return to vacuum solutions with $\mathbb{R}\times U(1)^2$ isometry group. As discussed above, the metric takes the canonical form 
\begin{equation}\label{Weyl5d}
g = -H \td t^2 + \frac{\lambda'_{ij}}{H^{1/2}}(\td\phi^i - w^i \td t)(\td\phi^j - w^j \td t) + e^{2\nu} (\td\rho^2 + \td z^2)
\end{equation} where $\rho^2 = \det \lambda'$ is harmonic on the orbit space.  Remarkably, the vacuum field equations for this spacetime can be derived from the critical points of the following functional, as first discussed by Carter for $D=4$ in \cite{carter1973black} (see \cite{hollands2008uniqueness} for general dimension):
\begin{equation}
\mathcal{M}' = \frac{\pi}{16} \int_{\tilde{\mathcal{B}}} \Tr \left(\mathcal{V}^{-1} \td \mathcal{V}\right)^2 \, \rho \td \rho \td z
\end{equation} where ${\tilde{\mathcal{B}}}$ is the orbit space of \emph{spaetime}, $\mathcal{V}$ is the $3 \times 3$ unimodular matrix
\begin{equation}
\mathcal{V}=\begin{pmatrix}
  \frac{1}{\det\lambda} & -\frac{Y_i}{\det\lambda} \\
  &\\
 -\frac{Y_j}{\det\lambda} & \lambda_{ij}+\frac{Y_iY_j}{\det\lambda}
 \end{pmatrix}
 \end{equation} where
 \begin{equation}\label{L1}
 \lambda_{ij} = \frac{\lambda'_{ij}}{H^{1/2}}  \end{equation} and $Y$ are the \emph{spacetime} twist potentials.  Note that it follows that $H = \rho^2 (\det \lambda)^{-1}$. 
That is, the Euler-Lagrange equations for $\mathcal{M'}$ are precisely those for the vacuum field equations for the above form of the metric. Once $\Phi$ is determined,  the remaining functions $H$ and conformal factor $\nu$ are determined by quadrature. 

Expanding out the Lagrangian gives
\begin{eqnarray}
\text{Tr}\left[\left(\mathcal{V}^{-1}\td\mathcal{V}\right)^2\right]
&=&\left(\frac{\td\det\lambda}{\det\lambda}\right)^2+\text{Tr}\left[\left(\lambda^{-1}\td\lambda\right)^2\right]+2\frac{\text{Tr}\left(\lambda^{-1}\td Y\td Y^t\right)}{\det\lambda}\label{1}
\end{eqnarray} We wish to express the action in terms of $\lambda'_{ij}$. Since
\begin{equation}
\td\lambda=\frac{1}{2}\left(\frac{\det\lambda}{\rho^2}\right)^{-\frac{1}{2}}\left(\frac{\td\det\lambda}{\rho^2}-2\frac{\det\lambda\td\rho}{\rho^3}\right)\lambda'+\left(\frac{\det\lambda}{\rho^2}\right)^{\frac{1}{2}}\td\lambda' \;,
\end{equation} a calculation yields
\begin{eqnarray}
\text{Tr}\left[\left(\lambda^{-1}\td\lambda\right)^2\right]
&=&\frac{1}{2}\left(\frac{\td\det\lambda}{\det\lambda}\right)^2-2\left(\frac{\td\rho\cdot\td\rho}{\rho^2}\right)+\text{Tr}\left[\left(\lambda'^{-1}\td\lambda'\right)^2\right]\nonumber \label{2} \; .
\end{eqnarray} Note $\td \rho \cdot \td \rho = 1$. 

Consider a constant-time spatial slice of the stationary, axisymmetric metric \eqref{Weyl5d}. The metric can be placed in our general form for our initial data provided
\begin{equation}
\Phi^2 = e^{2v} = \frac{1}{H^{1/2}} = \left[ \frac{\det \lambda}{\det \lambda'}\right]^{1/2}
\end{equation} which implies
\begin{equation}\label{L2}
v = \frac{1}{4} \log (\det \lambda) - \frac{\log \rho}{2} \; .
\end{equation} We then deduce
\begin{eqnarray}
\td v \cdot \td v &=&\left(\frac{\td\det\lambda}{4\det\lambda}-\frac{\td\rho}{2\rho}\right)^2 =\frac{1}{16}\left(\frac{\td\det\lambda}{\det\lambda}\right)^2-\frac{1}{4}\td\left(\log\rho\right)\cdot\td\log\left(\frac{\rho}{\det\lambda}\right)\label{7} \; .
\end{eqnarray} Using equations \eqref{L1} and \eqref{L2} one can  replace the independent variables $v$ and $\lambda'_{ij}$ by $\det{\lambda}$ and $\lambda_{ij}$ in the mass functional. Then $\det{\lambda}$, $\lambda_{ij}$, and $Y^i$ are taken to be independent, and we have \footnote{Precisely, this equality holds only if the integration domain $\Omega \subset \tilde{\mathcal{B}}$ does not include the axis $\rho =0$.}
\begin{eqnarray}
\mathcal{M}_{\tilde{\mathcal{B}}} &=& \mathcal{M}' + \frac{\pi}{4} \int_{\tilde{\mathcal{B}}} \left[ \frac{\td \rho \cdot \td \rho}{\rho} - \frac{3}{2} \frac{ \td \rho \cdot \td \det \lambda}{\det \lambda} \right] \td \rho \wedge \td z + \frac{\pi}{2}\int_{\partial {\tilde{\mathcal{B}}}} \alpha \nonumber \\
&=& \mathcal{M}'  + \frac{\pi}{4} \int_{\tilde{\mathcal{B}}} \td \left[\log \left(\frac{\rho}{(\det \lambda)^{3/2}}\right) \td z \right] + \frac{\pi}{2}\int_{\partial \tilde{\mathcal{B}}} \alpha \nonumber \\
&=& \mathcal{M}'   + \frac{\pi}{4}\int_{\partial \tilde{\mathcal{B}} \cup \tilde{\mathcal{B}}_\infty}\left( 2\alpha + \log \left(\frac{\rho}{(\det \lambda)^{3/2}}\right) \td z \right) 
\end{eqnarray} where $\alpha$ is the one-form defined in \eqref{alpha}. Note that we have taken the domain of integration in $\mathcal{M}$ to be over $\tilde{\mathcal{B}}$ when demonstrating this equivalence.  This is an important point, because $\partial \tilde{\mathcal{B}}$ will, in general, contain additional finite timelike rods on the axis $\rho =0$ corresponding to Killing horizons (i.e. where a timelike Killing vector filed becomes null) which are not present on $\partial \mathcal{B}$.  The domain of integration of $\mathcal{M}'$ covers only the exterior region to the black hole, with an inner boundary representing the horizon. In contrast, our mass functional is naturally defined over $\mathcal{B}$ and covers a complete manifold with no inner boundary, and in particular may have additional asymptotic regions.  In general, $\mathcal{M}_{\tilde{\mathcal{B}}}$ will be singular because it may diverge on the horizon rod, whereas, $\mathcal{M}$ is finite. In the special case of \emph{extreme} horizons, however, the orbit spaces coincide, because the timelike horizon rod shrinks to a point and corresponds to an asymptotically cylindrical region \cite{Figueras2010}.  

In summary, we have shown that over an appropriate domain, $\mathcal{M}$ equals Carter's functional, up to a divergent boundary term. Equivalently, we have proved that if one considers the change of variables $(v, \lambda'_{ij}) \to \lambda_{ij}$ given by \eqref{L1} and \eqref{L2},  then $\mathcal{M}$ is precisely the same as $\mathcal{M'}$ up to a boundary term.  It follows they have the same Euler-Lagrange equations, provided we consider variations which are fixed on $\partial \tilde{\mathcal{B}}$.  Hence the critical points of Carter's functional. i.e. the stationary, axisymmetric vacuum solutions, are also critical points of the mass functional. 

It is interesting to directly compute the Euler-Lagrange equations of $\mathcal{M}$.  The details are tedious and we simply summarize the result here.  First we vary the mass functional with respect to functions on the orbit space $\bar{\lambda}'$, $\bar{v}$ and $\bar{Y}$, which have compact supported on the interior of $\mathcal{B}$ (and in particular vanish on $\partial{\mathcal{B}}$ and $\mathcal{B}_\infty$).
 Therefore, by \cite{alaee2014small} the angular momenta will be preserved. We define
\begin{equation}
\mathcal{E}(\epsilon)=\mathcal{M}\left(v+\epsilon\bar{v},\lambda'+\epsilon\bar{\lambda'},Y+\epsilon\bar{Y}\right)
\end{equation}
Then we have
\begin{eqnarray}
\delta\mathcal{E}(0)&=&\frac{\pi}{4}\int_{\mathcal{B}}\Bigg[12\td v \cdot \td\bar{v} -\frac{1}{2\rho^2}\left(\td\bar{\lambda'}_{11}\cdot\td\lambda'_{22}+\td\lambda'_{11}\cdot\td\bar{\lambda'}_{22}
-2\td\bar{\lambda'}_{12}\cdot\td\lambda'_{12}\right) \\
&+&e^{-6v}\left((\det\bar{\lambda'})\frac{\text{Tr}(\bar{\lambda'}^{-1}\td Y\td Y^t)}{2\rho^4}+\frac{\text{Tr}({\lambda'}^{-1}\td\bar{Y}\td Y^t)}{\rho^2}-3\bar{v}\frac{\text{Tr}({\lambda'}^{-1}\td{Y}\td Y^t)}{\rho^2}\right)
\Bigg]\, \rho \td \rho \td z \nonumber
\end{eqnarray} Performing the appropriate integration by parts and imposing $\delta\mathcal{E}(0) = 0$ yields 
\begin{eqnarray}
4\Delta_3v+e^{-6v}\frac{\text{Tr}\left({\lambda'}^{-1}\nabla{Y} \cdot \nabla Y^t\right)}{\rho^2} &=
&0 \; , \label{EL1}\\
\nabla\left(\frac{\lambda'}{\rho^2}\right)+\frac{e^{-6v}}{\rho^4}\nabla Y \cdot \nabla Y^t&=&0 \; ,\qquad
\nabla\left(\frac{e^{-6v}}{\rho^2}\lambda'^{-1}\nabla Y\right)=0 \end{eqnarray} where $\Delta_3$ is the Laplacian with respect to the metric \eqref{3flat}. 
Consider the critical points of $\mathcal{M}$ that are extreme, stationary, axisymmetric vacuum solutions.  We can use the above to show that the mass functional is positive definite for these data. We have $h=e^{2v}\tilde{h}$  and thus
\begin{equation}\label{scalarcurvh}
\tilde{R}=Re^{2v}+6\Delta_{\tilde{h}} v+6e^{-2U}(\td v)^2
\end{equation}
Since $\Delta_{\tilde{h}}=e^{-2U}\Delta_3$ on $U(1)^2$-invariant functions, using \eqref{eq:constraints}, \eqref{tildeK^2}  and \eqref{EL1} yields\footnote{Note that the coefficient $-3$ is given incorrectly in the journal version.}
\begin{eqnarray}
\tilde{R}
&=&e^{-2U}\left(-3e^{-6v}\frac{\text{Tr}\left({\lambda'}^{-1}\td{Y}\td Y^t\right)}{2\rho^2}+6(\td v)^2\right)
\end{eqnarray} Substitution into the expression \eqref{m} gives
\begin{equation}\label{massextreme}
\mathcal{M}_{\text{cp}}=\frac{3\pi}{4}\int_{\mathcal{B}}e^{-6v}\frac{\text{Tr}\left(\lambda'^{-1}\td Y\td Y^t\right)}{2\det\lambda'}\, \,\rho \td\rho \td z 
\end{equation} where $\mathcal{M}_{\text{cp}}$ is the restriction of $\mathcal{M}$  to these critical points. Clearly this is positive definite.
\section{Positivity of $\mathcal{M}$}\label{positivity}
In this section we investigate the positivity of $\mathcal{M}$. Positivity is a desirable property as it plays a key role in applications to geometric inequalities for three-dimensional initial data \cite{dain2006proof,dain2008proof,dain2008inequality} and investigating the linear stability of extreme black holes\footnote{The stability argument uses positivity of the second variation of the mass functional about extreme Kerr. This energy is related to the recent construction of Hollands-Wald \cite{Hollands:2012sf} of a canonical energy, which has recently been used to demonstrate the existence of instabilities of (near)-extreme black holes \cite{Hollands:2014lra}. } \cite{Dain:2014iba}.  Gibbons and Holzegel \cite{gibbons2006positive} have generalized Brill's proof of positive mass for a restricted class of four-dimensional initial data with $U(1)^2$ isometry, by expressing the mass in a manifestly positive definite way.  We will show that for a particular set of initial data, $\mathcal{M}$ can be expressed in a form such that the arguments of \cite{gibbons2006positive} can be adapted to demonstrate positivity. It is important to note that our boundary conditions are weaker than the ones used in \cite{gibbons2006positive}.  A proof of positivity for arbitrary rod data remains to be found. In the following, we will consider  asymptotically flat data with a single additional asymptotic end with $N \cong S^3$. 

\begin{figure}[h]
\centering
\subfloat[{Orbit space as half plane}]{
\begin{tikzpicture}[scale=.85, every node/.style={scale=0.6}]
\fill[ fill=black!50!blue, opacity=0.35,very thick](-4,4)--(-4,0)--(4,0)--(4,4);
\draw[black,ultra thick](-4,0)--(4.3,0)node[black,font=\large,right=.2cm]{$z$};
\draw[black,thick](0,0)--(0,4.3)node[black,font=\large,above=.2cm]{$\rho$};
\draw[red,fill=red] (-3,0) circle [radius=.07] node[black,below=.1cm]{$a_1$};
\draw[red,fill=red] (-2,0) circle [radius=.07] node[black,below=.1cm]{$a_2$};
\draw[red,fill=red] (-.8,0) circle [radius=.07] node[black,below=.1cm]{$a_{i-1}$};
\draw[fill=white] (0,0) circle [radius=.08] node[black,below=.1cm]{$a_E$};
\draw[red,fill=red] (.8,0) circle [radius=.07] node[black,below=.1cm]{$a_{i+1}$};
\draw[red,fill=red] (1.6,0) circle [radius=.07] node[black,below=.1cm]{$a_{i+2}$};
\draw[red,fill=red] (3,0) circle [radius=.07] node[black,below=.1cm]{$a_{n}$};
\end{tikzpicture}}
\hspace{3em}
\subfloat[{ Orbit space as infinite strip }]{
\begin{tikzpicture}[scale=.85, every node/.style={scale=0.6}]
\draw[black,thick](-3,0)--(3.3,0)node[black,font=\large,right=.2cm]{$y$};
\draw[black,thick](0,-2)--(0,2)node[black,font=\large,above=.2cm]{$x$};
\fill[ black!50!blue, opacity=0.35](-3,1)--(3,1)--(3,-1)--(-3,-1);
\draw[black,ultra thick](-3,1)--(3,1)node[black,font=\large,right=.2cm]{$\mathcal{I}^+$};
\draw[black,ultra thick](-3,-1)--(3,-1)node[black,font=\large,right=.2cm]{$\mathcal{I}^-$};
\draw[red,fill=red] (-2.4,-1) circle [radius=.07] node[black,below=.1cm]{$a_{i-1}$};
\draw[red,fill=red] (-.7,-1) circle [radius=.07] node[black,below=.1cm]{$a_{i-2}$};
\draw[red,fill=red] (.7,-1) circle [radius=.07] node[black,below=.1cm]{$a_{2}$};
\draw[red,fill=red] (1.6,-1) circle [radius=.07] node[black,below=.1cm]{$a_{1}$};
\draw[red,fill=red] (-2.2,1) circle [radius=.07] node[black,above=.1cm]{$a_{i+1}$};
\draw[red,fill=red] (-1.3,1) circle [radius=.07] node[black,above=.1cm]{$a_{i+2}$};
\draw[red,fill=red] (.5,1) circle [radius=.07] node[black,above=.1cm]{$a_{n-1}$};
\draw[red,fill=red] (2,1) circle [radius=.07] node[black,above=.1cm]{$a_{n}$};
\end{tikzpicture}}
\caption{The rod point $a_E=\{\rho,z=0\}$ represents another end. (a) and (b) illustrates the map from the $z+i\rho$ complex plane to the $y+ix$ complex plane where $y=\log r$.}\label{fig4}
\end{figure}
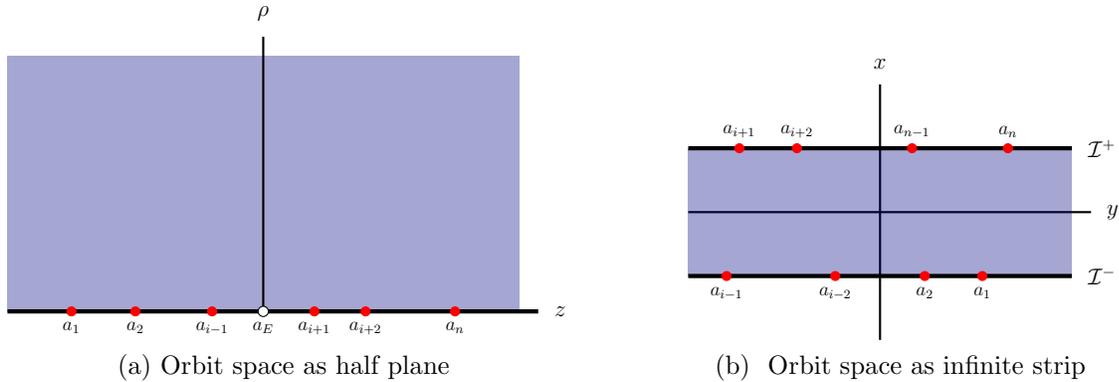

 Introduce the coordinates $(r,x)$ given by the transformation \eqref{rx}. This is equivalent to introducing a map from $\mathcal{B}\cong\mathbb{R}\times\mathbb{R}^+\backslash \{a_E\}$ to the infinite strip $\mathcal{B}\cong\mathbb{R}\times [-1,1]$ \cite{Figueras2010}. This map will divide the axis $\rho=0$ into two disconnected  axes $\mathcal{I}^{\pm}=\{r>0,\,x=\pm 1\}$ and another end,  $a_E=\{r=0,\, |x|\leq 1\}$. Note that  the rod structure is contained on $\mathcal{I}^{\pm}$,  see Figure \ref{fig4}. 

Consider the mass functional \eqref{mass3d} and \eqref{mass}.  The inner products are taken with respect to $\delta_2=\td\rho^2+\td z=r^2(\td r^2+\frac{r^2\td x^2}{4(1-x^2)})=r^2\delta'_2$.  We will rewrite the functional with respect to $\delta'_2$ and as an integral over the infinite strip. Thus we have 
\begin{eqnarray} \label{massrx}
\mathcal{M}=\frac{1}{32}\int_{\mathcal{B}}\Bigg(-\frac{\det\td\lambda'}{2\det\lambda'}+ e^{-6v}\frac{\text{Tr}\left(\lambda'^{-1}\td Y \cdot \td Y^t\right)}{2\det\lambda'}+6\left(\td v\right)^2\Bigg)\,  r^3\td r\td x+\frac{1}{4}\int_{\partial\mathcal{B}\cup\mathcal{B}_{\infty}}\alpha
\end{eqnarray}
where  all scalar products of one-forms are taken with respect to $\delta'_2$. Note that $(\rho,z)$ and $(x,r)$ have positive orientation. The boundary is $\partial\mathcal{B}\cup \mathcal{B}_{\infty}=\mathcal{I}_E+\mathcal{I}_{\infty}+\mathcal{I}^++\mathcal{I}^-$ where $\mathcal{I}_E\equiv\{r=0,-1\leq x\leq 1\}$, $\mathcal{I}_{\infty}\equiv\{r=\infty,-1\leq x\leq 1\}$. In terms of the $(r,x)$ chart, we have 
\begin{figure}[h]
\centering
\subfloat[{Orbit space as half plane}]{
\begin{tikzpicture}[scale=.85, every node/.style={scale=0.6}]
\fill[ fill=black!50!blue, opacity=0.35,very thick](-4,4)--(-4,0)--(4,0)--(4,4);
\draw[black,ultra thick](-4,0)--(4.3,0)node[black,font=\large,right=.2cm]{$z$};
\draw[black,thick](0,0)--(0,4.3)node[black,font=\large,above=.2cm]{$\rho$};
\draw[red,fill=red] (-3.6,0) circle [radius=.07] node[black,below=.1cm]{$a_1$};
\draw[red,fill=red] (-2.3,0) circle [radius=.07] node[black,below=.1cm]{$a_2$};
\draw[red,fill=red] (-.6,0) circle [radius=.07] node[black,below=.1cm]{$a_{3}$};
\draw[fill=white] (0,0) circle [radius=.08] node[black,below=.1cm]{$a_E$};
\draw[red,fill=red] (1,0) circle [radius=.07] node[black,below=.1cm]{$a_{5}$};
\draw[red,fill=red] (1.6,0) circle [radius=.07] node[black,below=.1cm]{$a_{6}$};
\draw[red,fill=red] (3,0) circle [radius=.07] node[black,below=.1cm]{$a_{7}$};
\draw[blue] (3,0) arc (0:180:3cm);
\draw[blue] (3.6,0) arc (0:180:3.6cm);
\draw[blue] (2.3,0) arc (0:180:2.3cm);
\draw[blue](.6,0) arc (0:180:.6cm);
\draw[blue] (1,0) arc (0:180:1cm);
\draw[blue](1.6,0) arc (0:180:1.6cm);
\end{tikzpicture}}
\hspace{3em}
\subfloat[{ Orbit space as infinite strip }]{
\begin{tikzpicture}[scale=.85, every node/.style={scale=0.6}]
\draw[black,thick](-3,0)--(3.3,0)node[black,font=\large,right=.2cm]{$y$};
\draw[black,thick](0,-2)--(0,2)node[black,font=\large,above=.2cm]{$x$};
\fill[black!50!blue, opacity=0.35](-3,1)--(3,1)--(3,-1)--(-3,-1);
\draw[black,ultra thick](-3,1)--(3,1)node[black,font=\large,right=.2cm]{$\mathcal{I}^+$};
\draw[black,ultra thick](-3,-1)--(3,-1)node[black,font=\large,right=.2cm]{$\mathcal{I}^-$};
\draw[red,fill=red] (-2.4,-1) circle [radius=.07] node[black,below=.1cm]{$a_{3}=b_1$};
\draw[red,fill=red] (.5,-1) circle [radius=.07] node[black,below=.1cm]{$a_{2}=b_4$};
\draw[red,fill=red] (2.3,-1) circle [radius=.07] node[black,below=.1cm]{$a_{1}=b_6$};
\draw[red,fill=red] (-1.7,1) circle [radius=.07] node[black,above=.1cm]{$a_{5}=b_2$};
\draw[red,fill=red] (-.5,1) circle [radius=.07] node[black,above=.1cm]{$a_{6}=b_3$};
\draw[red,fill=red] (1.5,1) circle [radius=.07] node[black,above=.1cm]{$a_{7}=b_5$};
\draw[blue](-2.4,-1)--(-2.4,1)node[black,below=1cm,left=.1cm]{$A_{0}$};
\draw[blue](.5,-1)--(.5,1)node[black,below=1cm,left=.01cm]{$A_{3}$};
\draw[blue](2.3,-1)--(2.3,1)node[black,below=1cm,left=.3cm]{$A_{5}$}node[black,below=1cm,right=.04cm]{$A_{6}$};
\draw[blue](-1.7,-1)--(-1.7,1)node[black,below=1cm,left=.1cm]{$A_{1}$};
\draw[blue](-.5,-1)--(-.5,1)node[black,below=1cm,left=.5cm]{$A_{2}$};
\draw[blue](1.5,-1)--(1.5,1)node[black,below=1cm,left=.5cm]{$A_{4}$};
\end{tikzpicture}}
\caption{The orbit space can be subdivided into subregions $A_i$ which are half-annuli in the $(\rho,z)$ plane and  rectangles in the $(y,x)$ plane. In this case $n=7$.}\label{fig5}
\end{figure}
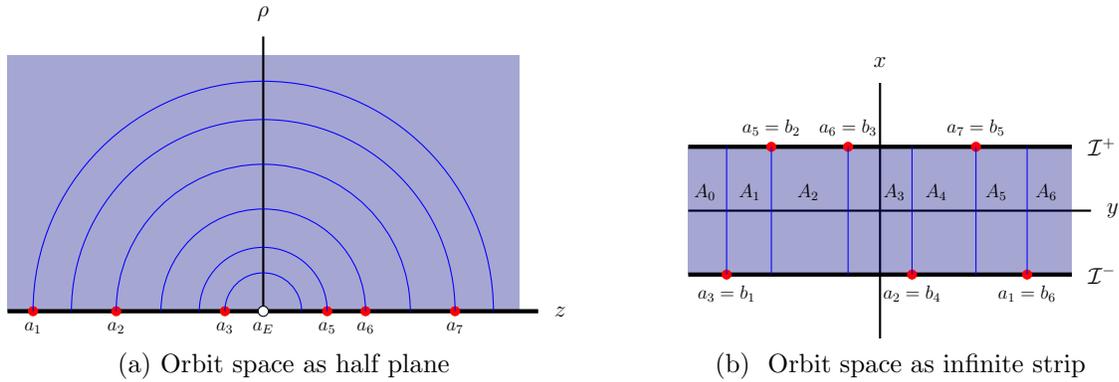
\begin{eqnarray}
\alpha =-\left(r(1-x^2)V_{,x}+rxV\right)\td r+
\left(\frac{r^3}{4}V_{,r}-\frac{r^2}{2}V\right)\td x
\end{eqnarray}
Then we have\footnote{The condition on $\alpha|_{\mathcal{I}_E}$ in the journal version is related to $V$ evaluated on the end and is modified here.}
\begin{eqnarray}
\alpha|_{\mathcal{I}_E}= 0,\qquad\alpha|_{\mathcal{I}_{\infty}}=-\bar{V}(x)\td x,\qquad\alpha|_{\mathcal{I}^+}=-rV|_{x=1}\td r,\qquad\alpha|_{\mathcal{I}^-}=rV|_{x=-1}\td r\nonumber
\end{eqnarray}
Thus with appropriate orientation
\begin{eqnarray}\label{alpharx}
\int_{\partial\mathcal{B}\cup \mathcal{B}_{\infty}}\alpha
=\int_{0}^{\infty}r\left(V|_{x=1}+V|_{x=-1}\right)\td r = \sum_{i=0}^{n-1}\int_{b_i}^{b_{i+1}}r\left(V|_{x=1}+V|_{x=-1}\right)\td r
\end{eqnarray} 

Consider the integral \eqref{massrx}.  We are given $n$ rod points $a_i$. Subdivide the infinite strip into $n$ rectangular columns $A_i$ with
\begin{equation}
A_i  = \left\{ -1 \leq x \leq 1, b_i < r < b_{i+1} \right \}\;,  \quad i=0\ldots n-1
\end{equation} where $b_i$ correspond to the location of the rod points $a_i$ after ordering along the $y = \log r$ axis (see Figure \ref{fig5}). For convenience, we have chosen $b_1 < b_2 < \ldots  < b_{n-1}$ We take $b_0 = 0$ to correspond to the asymptotic end $a_E$ and $b_n$ to correspond to the asymptotically flat end $r \to \infty$.  We then express \eqref{massrx} as
\begin{equation}
\mathcal{M} = \sum_{i=0}^{n-1} \int_{A_i} \mathcal{M}_i 
\end{equation} where $\mathcal{M}_i$ is the restriction of $\mathcal{M}$ to $A_i$. \\ \par \noindent
Fix a region $A_i$.  Then one of the following two possibilities must occur: (a) distinct Killing fields $v_{(i)}$ and $w_{(i)}$ vanish on $A_i \cap \mathcal{I}^+$  and $A_i \cap \mathcal{I}^-$  respectively (in this case $A_i$ is topologically $S^3 \times \mathbb{R}$), or (b) the same Killing field $v_{(i)} = v_{(i)}^i m_i $ vanishes on both of the disjoint sub intervals  $A_i \cap \mathcal{I}^\pm$ (in this case $A_i$ is topologically $S^2 \times D$ where $D$ is a non-contractible disc) . We can demonstrate positivity for case (a).  In this case without loss of generality we can select the following parameterization of the 3 independent functions contained in $\lambda'_{ij}$ and $v$:
\begin{eqnarray}
\begin{aligned}\label{lambdaGH}
\lambda'_{11}&=\frac{r^2(1-x)}{2\sqrt{1-W^2}}e^{V_1-V_2}\qquad
\lambda'_{22}=\frac{r^2(1+x)}{2\sqrt{1-W^2}}e^{V_2-V_1}\\
\lambda'_{12}&=\frac{r^2\sqrt{1-x^2}W}{2\sqrt{1-W^2}}\qquad v=\frac{V_1+V_2+\log\sqrt{1-W^2}}{2}
\end{aligned} 
\end{eqnarray} where without loss of generality we have chosen $v_{(i)} = \partial_{\phi_1}$ and  $w_{(i)} = \partial_{\phi_2}$. $V_1,V_2$ and $W$ are $C^1$ functions whose boundary conditions on the axis are induced from those of $\lambda'_{ij}$ and $v$ \eqref{Vinfty} and \eqref{lambdainfty}.  In particular, we have $\det\lambda'=\rho^2$ and to remove conical singularities on $\mathcal{I}^{\pm}$  \eqref{regaxis} we require:
\begin{equation}
2V-V_1+V_2= 0 \quad \text{on $\mathcal{I}^{+}$},\quad 2V-V_2+V_1= 0 \quad \text{on $\mathcal{I}^{-}$},\quad W=0\quad \text{on $\mathcal{I}^{\pm}$}\label{GR1}
\end{equation} Note that since $\lambda'_{ij}$ and $v$ are continuous across the boundary of $A_i$, this will impose boundary conditions on the parameterization functions in adjacent subregions. 
Secondly, we rewrite the second and fourth terms of $\mathcal{M}$ as functions of $V_1$, $V_2$, and $W$, yielding:
\begin{eqnarray}\label{RicciGH1}
&&\frac{\det d\lambda'}{2\det\lambda'}= \\ &\phantom{=}& \frac{-1}{2(1-W^2)}\Bigg[(\td V_1-\td V_2)^2-\frac{8}{r^2}\partial_{x}(V_1-V_2)+(\td W)^2+\frac{W^2(\td W)^2}{1-W^2} 
+\frac{4W^2}{r^2(1-x^2)}\Bigg] \nonumber
\end{eqnarray}
and
\begin{eqnarray}\label{RicciGH2}
6(\td v)^2&=&\frac{3}{2}(\td V_1+\td V_2)^2+\frac{3}{2}\frac{W^2(\td W)^2}{(1-W^2)^2}-\frac{3W}{1-W^2}(\td V_1\cdot\td W+\td V_2\cdot\td W)
\end{eqnarray}
Therefore, we have 
{\small\begin{eqnarray}
\mathcal{M}_i
&=&\frac{1}{32}\int_{A_i}\Bigg(Re^{2v + 2U}+(\td V_1+\td V_2)^2+(\td V_1)^2+(\td V_2)^2\\
&+&\frac{W^2}{2(1-W^2)}\left[(\td V_1-\td V_2)^2-\frac{6}{W}(\td V_1\cdot\td W+\td V_2\cdot\td W)\right]\nonumber\\
&+&\frac{W^2}{r^2(1-W^2)}\left[4\partial_xV_2-4\partial_xV_1+\frac{2}{(1-x^2)}\right]+\frac{(\td W)^2}{2(1-W^2)}+\frac{2W^2(\td W)^2}{(1-W^2)^2}\Bigg)\,r^3 \td x\td r\nonumber\\
&+&\frac{1}{8}\int_{b_i}^{b_{i+1}}r\left((V_1-V_2)|_{x=-1}-(V_1-V_2)|_{x=1}\right)\td r+\frac{1}{4}\int_{b_i}^{b_{i+1}}r\left(V|_{x=1}+V|_{x=-1}\right)\td r\nonumber\\
&=&\frac{1}{32}\int_{A_i}\Bigg(Re^{2v + 2U}+(\td V_1+\td V_2)^2+(\td V_1)^2+(\td V_2)^2\nonumber\\
&+&\frac{W^2}{2(1-W^2)}\left[(\td V_1-\td V_2)^2-\frac{6}{W}(\td V_1\cdot\td W+\td V_2\cdot\td W)\right]\nonumber\\
&+&\frac{W^2}{r^2(1-W^2)}\left[4\partial_xV_2-4\partial_xV_1+\frac{2}{(1-x^2)}\right]+\frac{(\td W)^2}{2(1-W^2)}+\frac{2W^2(\td W)^2}{(1-W^2)^2}\Bigg)\,r^3 \td x\td r\nonumber
\end{eqnarray}} Consider the first equality. The first term follows from the constraint equation for maximal slices, \eqref{tildeK^2}, and \eqref{conformaldata}. The remaining bulk terms follow from  \eqref{RicciGH1} and \eqref{RicciGH2} while the first boundary term comes from \eqref{RicciGH1}  and the second from \eqref{alpharx} .  The second equality is obtained by noting the boundary contributions cancel by regularity on the axes \eqref{GR1}. The remaining terms  can be shown to be positive by a straightforward application of the arguments given in section 4.3 of \cite{gibbons2006positive}. Therefore, $\mathcal{M}_i \geq 0$.

From this result it follows that provided \emph{all} subregions $A_i$ fall into class (a) then $\mathcal{M}$ is positive-definite.  In particular, for the rod structure of Myers-Perry initial data, there is only one region $A_0$ of class (a) and hence for any data with the same rod structure, $\mathcal{M} \geq 0$.  One might expect a similar argument to hold for class (b). This case of course includes initial data for black rings (the same Killing vector field vanishes on either side of the asymptotic end).  By choosing a general parametrization for the various functions in this region, one finds that the boundary term has an indefinite sign.  However our strategy is merely sufficient to demonstrate positivity, and we expect positivity will hold for general rod structure. Interestingly, for the initial data for extreme black rings, the expression \eqref{massextreme} shows $\mathcal{M} \geq 0$.

\section{Discussion}
We have constructed a mass functional $\mathcal{M}$ valid for a broad class of asymptotically flat $t-\phi^i$-symmetric maximal initial data for the vacuum Einstein equations in five dimensions. $\mathcal{M}$ can be considered an extension of a similar functional defined for three-dimensional initial data sets \cite{dain2006variational} .  We can check this mass functional is finite and evaluates to the ADM mass provided certain boundary and asymptotic conditions are met.  These conditions encompass a large class of initial data, and in particular we have checked this explicitly for the usual maximal constant-time slices for the Myers-Perry black hole (see Appendix \ref{appA}) and the extreme vacuum black ring solution. Moreover, we proved that $\mathbb{R} \times U(1)^2$-invariant solutions of the vacuum Einstein equations are critical points of this functional amongst this class of data.  Finally, we have shown explicitly that the mass functional is positive for a particular class of rod structures as explained in detail above, although it remains to show this for an arbitrary rod data. This property is relevant to investigate geometric inequalities for five-dimensional vacuum solutions. An starting towards this goal is to show a \emph{local} mass-angular momenta inequality along the lines of \cite{dain2006proof}. This problem is currently under investigation.

\section{Acknowledgments}
We would like to thank S Hollands for clarifying the relationship of mass functionals and his recent construction with R Wald of a canonical energy.  HKK also thanks J Lucietti for comments concerning uniqueness theorems for extreme black holes. AA is partially supported by a graduate scholarship from Memorial University. HKK is supported by an NSERC Discovery Grant.

\appendix
\numberwithin{equation}{section}
\section{Mass of conformal metric}\label{appB}
Assume that we have an asymptotically flat initial data set $(\Sigma,h_{ab},K_{ab})$ of Einstein's equation.The ADM mass of this data is given by formula \eqref{ADMmass}. But by a rescaling similar to \eqref{conformaldata}  we have
\begin{equation}
M_{ADM}=-\frac{3}{8\pi}\lim_{r\rightarrow\infty}\int_{S^3_r}n^c \tilde{\nabla}_c\Phi \, \td s_{\tilde{h}}+\tilde{M}_{ADM}
\end{equation}
where $\tilde{M}_{ADM}$ is the ADM mass of $\tilde{h}$. Now as in Section \ref{positivity} we can introduce a chart with coordinates $(r,x)$ such that the asymptotically flat conformal metric takes the form
\begin{equation}\label{AFtildeh}
\tilde{h}=e^{2V}\left(\td r^2+\frac{r^2}{4(1-x^2)} \td x^2\right)+f_2\frac{r^2}{2}(1-x)\td\phi^2+f_3\frac{r^2}{2}(1+x)\td\psi^2+f_4r^2(1-x^2)\td\phi\td\psi
\end{equation}
with the fall-off conditions $e^{2V}-1, f_2-1$, and  $f_3-1=\mathcal{O}(r^{-2})$ and $f_4=o(r^{-2})$ as  $r \to \infty $. 
Then the ADM mass of the conformal metric is
\begin{eqnarray}
\tilde{M}_{ADM}&=&-\frac{1}{16\pi}\lim_{r\rightarrow\infty}\int_{S^3}\left(r^2 \partial_r\left[ r(f_2+f_3-2)\right] + r^5\partial_r\left(\frac{e^{2V}-1}{r^2}\right)\right)\,\td\Omega_3\nonumber\\
&=&\frac{1}{16\pi}\int_{S^3}\left(f(x)+g(x)\right)\,\td\Omega_3+\frac{1}{2\pi}\int_{S^3}\bar{V}(x) \,\td\Omega_3\nonumber\\
&=& \frac{\pi}{2}\int_{-1}^{1} \bar{V}(x) \, \td x \label{Vbar}
\end{eqnarray}  
The first equality is the definition of ADM mass applied to \eqref{AFtildeh}. The second equality uses the expansion of $\lambda'_{ij}$ and $V$ at infinity \eqref{lambdainfty}, \eqref{Vinfty}.  Therefore, we can see the ADM mass of the conformal metric is zero if and only if the right hand side of  \eqref{Vbar} vanishes. It is trivially satisfied if $V = o(r^{-2})$. In general however, one may wish to consider weaker fall-off conditions on $V$ that still lead to vanishing ADM mass of the conformal metric.  In particular, we have checked explicitly for the general Myers-Perry black hole and for the extreme doubly spinning black ring that the right hand side of \eqref{Vinfty} vanishes, although $\bar{V}(x) \neq 0$ in these cases. 

\section{Myers-Perry initial data}\label{appA}
Here we consider the  Myers-Perry solution with coordinates $(t,\tilde{r},\theta,\phi_1,\phi_2)$ \cite{myers2011myers}.  The $\phi_i$ have period $2\pi$. Then we have following metric functions
\begin{eqnarray}
\omega^{1}&=&\frac{\mu a\lambda_{22} \sin^2\theta-\mu b\lambda_{12} \cos^2\theta}{\Sigma\det\lambda}\qquad
\omega^{2}=\frac{\mu b\lambda_{11} \cos^2\theta-\mu a\lambda_{12} \sin^2\theta}{\Sigma\det\lambda}\\
\lambda_{11}&=&\frac{a^2\mu}{\Sigma}\sin^4\theta+(\tilde{r}^2+a^2)\sin^2\theta\qquad \lambda_{12}=\frac{ab\mu}{\Sigma}\sin^2\theta\cos^2\theta\\
\lambda_{22}&=&\frac{b^2\mu}{\Sigma}\cos^4\theta+(\tilde{r}^2+b^2)\cos^2\theta
\end{eqnarray}
where
\begin{eqnarray}
\Sigma&=&\tilde{r}^2+b^2\sin^2\theta+a^2\cos^2\theta,\\
\Delta(\tilde{r})&=&\left(\tilde{r}^2+a^2\right)\left(\tilde{r}^2+b^2\right)-\mu \tilde{r}^2.
\end{eqnarray}
The metric on a constant time slice will be 
\begin{eqnarray}
h=\frac{\Sigma}{\Delta(\tilde{r})}\td r^2
+\Sigma \td\theta^2+\lambda_{ij}\td\phi^i\td\phi^j
\end{eqnarray}
This metric is singular at two roots $\tilde{r}_{\pm}$ of $\Delta(\tilde{r})$ which correspond to spacetime inner and outer horizons.  One can define a quasi-isotropic coordinate as
\begin{eqnarray}
\tilde{r}^2=r^2+\frac{1}{2}\left(\mu-a^2-b^2\right)+\frac{\mu\left(\mu-2a^2-2b^2\right)+(a^2-b^2)^2}{16r^2}
\end{eqnarray}
Note the outer horizon at $\tilde{r}_+$ is shifted to $r=0$ and the slice metric will be 
\begin{eqnarray}
h=\frac{\Sigma}{r^2}\left(\td r^2
+ r^2 \td\theta^2\right)+\lambda_{ij}\td\phi^i\td\phi^j
\end{eqnarray}
where $0<r<\infty$, $0<\theta<\pi/2$, and $0<\phi_1,\phi_2<2\pi$. The point $r=0$ is another asymptotic infinity  (see figure \ref{fig1}) and one can show this with computing the distance to  $r =0$ along a curve of constant $(\theta,\phi_1,\phi_2)$ from $r = r_0$, i.e.
\begin{equation}
\text{Distance}=\int_{r}^{r_0}\frac{\sqrt{\Sigma}}{r}dr\rightarrow \infty \qquad \text{as} \quad r\to 0
\end{equation}
In the extreme limit $\mu=(a+b)^2$ the quasi-isotropic radius simplifies to \cite{alaee2014small}
\begin{equation}
\tilde{r}^2=r^2+ab
\end{equation} 
The conformal metric $\tilde{h}$ can be determined by the relations
\begin{equation}\label{htildeMP}
\Phi^2=\frac{\sqrt{\det\lambda}}{\rho}, \qquad e^{2U}=\frac{\rho\Sigma}{r^4\sqrt{\det\lambda}}, \qquad \lambda'_{ij}=\Phi^{-2}\lambda_{ij}
\end{equation}
where $\rho=\frac{1}{2}r^2\sin 2\theta$ and $z=\frac{1}{2}r^2\cos 2\theta$. The potentials in the general case are cumbersome, but in the extreme case simplify to
{\small
\begin{eqnarray}
Y^1&=&\frac{a(a^2-b^2)(r^2+ab+b^2)\cos^2\theta-r^2a(2a^2+2ab+r^2)}{(a-b)^2}+\frac{a(r^2+ab+a^2)^2(r^2+ab+b^2)}{\Sigma(a-b)^2}\nonumber\\
&&\\
Y^2&=&\frac{br^2((a+b)^2+r^2)-b(a^2-b^2)(r^2+ab+a^2)\cos^2\theta}{(a-b)^2}-\frac{b(r^2+ab+a^2)(r^2+ab+b^2)^2}{\Sigma(a-b)^2}\nonumber
\end{eqnarray}}
The expansion at infinity is
\begin{eqnarray}
Y^1&=&\frac{a^3(a+b)^2}{(a-b)^2}-\frac{4J_1}{\pi}\cos^2\theta(2-\cos^2\theta)+\mathcal{O}(r^{-2})\\
Y^2&=&-\frac{ab^2(a+b)^2}{(a-b)^2}-\frac{4J_2}{\pi}\cos^4\theta+\mathcal{O}(r^{-2})
\end{eqnarray} The asymptotic behaviour of the conformal factor at infinity is given by
\begin{eqnarray}
\Phi&=&1+\frac{\mu}{4r^2}+\mathcal{O}(r^{-4})\qquad r\to\infty
\end{eqnarray} The region $r\to 0$ corresponds to another asymptotic region.  In the non-extreme case, we have
\begin{eqnarray}
\Phi&=&\frac{\sqrt[4]{(\mu-(a+b)^2)^2(\mu-(a-b)^2)^2}}{4r^2}+\mathcal{O}(1),\quad \Phi_{,r}=\mathcal{O}(r^{-3})\quad r\to 0
\end{eqnarray} and it is easy to verify that  $\tilde{h}$ approaches the flat metric on $\mathbb{R}^4$. Hence this region is an asymptotically flat end.  In the extreme case, however, one can check that
\begin{eqnarray}
\Phi = \frac{ (ab (a+b)^3)^{1/4}}{(a \cos^2\theta + b \sin^2\theta) r} + O(r) \qquad r \to 0
\end{eqnarray} By examining the behaviour of the metric $h$, one can see that the asymptotic region $r \to 0$ is a cylindrical end. In fact, explicit computation of $U$ and $\lambda'_{i}$ shows that the conformal metric $\tilde{h}$ approaches the metric of a cone over an $S^3$ equipped with an inhomogeneous metric, 
\begin{equation}
\tilde{h} = \Omega^2 \left( dr^2 + r^2 \gamma\right)
\end{equation} where $\Omega  = \Omega(\theta) \neq 0$ and $\gamma$ is conformal to the inhomogeneous metric on cross-sections of the horizon of the extreme Myers-Perry black hole. 
 
The conformal factor in either case satisfies conditions of \eqref{coninf} , \eqref{conformalmass1}, and we have at the asymptotic ends
\begin{equation}
\frac{3}{8\pi}\lim_{r\rightarrow 0}\int_{S_r}\frac{\tilde{\nabla}_c\Phi}{\Phi}\, n^c\,ds_{\tilde{h}}=0
\end{equation} 
where $ds_{\tilde{h}}=r^3\sin\theta\cos\theta+\mathcal{O}(r^{6})$. Now, one can expand the function $V$ at infinity and at the origin. As we discussed before we only consider behaviour of $V$ near $\rho=0$.  We find
\begin{eqnarray}
V&=&\frac{(a^2-b^2)\cos 2\theta}{4r^2}+\mathcal{O}(r^{-4})\qquad r\to\infty\\
V_+&=&\frac{2z+a^2+ab}{\sqrt{4z^2+3a^2b^2+2a^2z+b^4+2b^2z+4abz+a^3b+3ab^3}}\qquad z\in I_+\\
V_-&=&\frac{-2z+b^2+ab}{\sqrt{a^4+3a^3b+3a^2b^2-2a^2z+ab^3-4abz-2b^2z+4z^2}}\qquad z\in I_-
\end{eqnarray}
Thus $V$ satisfies condition \eqref{Vinfty} and \eqref{originV}. In particular, we read off $\bar{V} = \tfrac{1}{4}(a^2 - b^2) x$ and hence from \eqref{Vbar} we see $\tilde{h}$  (see \eqref{htildeMP}) has vanishing ADM mass.  In addition, when $z\to\pm\infty$
we have $V_{\pm}\to 1$ and $V_{\pm}$ are bounded continuous functions on rods $I_{\pm}$. Therefore, they are integrable. Let us consider boundedness of other terms in the mass functional \eqref{mass}.  We will consider explicitly the non-extreme case so the end is asymptotically flat. First we have the following expansion for $v$ at origin and infinity
\begin{equation}
(\td v)^2=-\frac{\mu}{2r^5}+\mathcal{O}(r^{-7})\quad r\to\infty \qquad
(\td v)^2=-\frac{2}{r^3}+\mathcal{O}(r^{-1})\quad r\to 0
\end{equation}
since the volume element is $\rho\td\rho\td z=r^5\sin\theta\cos\theta\, \td r\td\theta$, $(dv)^2$ is bounded at origin and infinity. Now we consider term which related to scalar curvature in mass functional \eqref{mass}. We use identity \eqref{2Didentity} and we have
\begin{equation}
\frac{\det\td\lambda'}{\det\lambda'}=\mathcal{O}(r^{-8})\quad r\to\infty \qquad
\frac{\det\td\lambda'}{\det\lambda'}=\mathcal{O}(1)\quad r\to 0
\end{equation}
This is clearly bounded.  One can check numerically over a range of $(a,b)$ that $\det \td \lambda' < 0$ everywhere. The only term remaining is related to the full contraction of extrinsic curvature and we have 
\begin{equation}
\frac{\text{Tr}\left(\lambda'^{-1}dYdY^{t}\right)}{2\det\lambda'}=\mathcal{O}(r^{-10})\quad r\to\infty \qquad
e^{-6v}\frac{\text{Tr}\left(\lambda'^{-1}dYdY^{t}\right)}{2\det\lambda'}=\mathcal{O}(r^{2})\quad r\to 0
\end{equation}
Therefore, non-extreme Myers-Perry lies in the domain on which the mass functional \eqref{mass} is defined. By similar steps the same result holds for the extreme case. 
\bibliographystyle{unsrt}
\bibliographystyle{abbrv}  
\bibliography{masterfile}
       
\end{document}